\documentclass[a4paper,11pt]{extarticle}  

\usepackage{stylesheet}
\usepackage{float}

\usepackage[left=1in,top=1in,bottom=.65in,right=1in]{geometry}

\newcommand*\rot{\rotatebox{45}}

\usepackage[page]{appendix} 

\usepackage{lipsum} 

\title{Uncertainties within Weather Regime definitions for the Euro-Atlantic sector in ERA5 and CMIP6}

\author{%
\textbf{%
Lotte Hompes \orcidlink{0009-0006-2605-800X},\textcolor{Accent}{\textsuperscript{1,2}} %
Swinda Falkena \orcidlink{0000-0002-2317-9213},\textcolor{Accent}{\textsuperscript{2}} 
Laurens Stoop \orcidlink{0000-0003-2756-5653}\textcolor{Accent}{\textsuperscript{1,*}} %
}\\[0.5em]
\begin{small}
\textcolor{Accent}{\textsuperscript{1}}Advanced Analytics, TenneT TSO, Arnhem, the Netherlands\\[0.5em] 
\textcolor{Accent}{\textsuperscript{2}}Institute for Marine and Atmospheric research Utrecht, Utrecht University, the Netherlands \\[0.5em]
\textcolor{Accent}{\textsuperscript{*}}Corresponding Author: \textcolor{Accent}{laurens.stoop@tennet.eu} \\ \end{small}
}

\date{10-01-2025}

\begin{document}









\maketitle


\begin{abstract}
\noindent
Certain Weather Regimes (WR) are associated with a higher risk of energy shortages, i.e. Blocking regimes for European winters.
However, there are many uncertainties tied to the implementation of WRs and associated risks in the energy sector. 
Especially the impact of climate change is unknown.

We investigate these uncertainties by looking at three methodologically diverse Euro-Atlantic WR definitions.
We carry out a thorough validation of these methods and analyse their methodological and spatio-temporal sensitivity using ERA5 data.
Furthermore, we look into the suitability of CMIP6 models for WR based impact assessments.

Our sensitivity assessment showed that the persistence and occurrence of regimes are sensitive to small changes in the methodology. 
We show that the training period used has a very significant impact on the persistence and occurrence of the regimes found.
For both WR4 and WR7, this results in instability of the regime patterns.

All CMIP6 models investigated show instability of the regimes.
Meaning that the normalised distance between the CMIP6 model regimes and our baseline regimes exceeds 0.4 or are visually extremely dissimilar. 
Only the WR4 regimes clustered on historical CMIP6 model data consistently have a normalised distance to our baseline regimes smaller than 0.4 and are visually identifiable. 
The WR6 definition exceeds the normalised distance threshold for all investigated CMIP6 experiments.
Though all CMIP6 model experiments clustered with the WR7 definition have a normalised distance to the baseline regimes below 0.4, visual inspection of the regimes indicates instability.

Great caution should be taken when applying WR's in impact models for the energy sector, due to this large instability and uncertainties associated with WR definitions.

\end{abstract}

\pagenumbering{arabic} 
\setcounter{page}{1}


\section{Introduction}
Human well-being is at risk due to Earth's changing climate~\citep{ipcc6_wg2}.
To mitigate greenhouse gas emissions, and their adverse effects on our climate, society is shifting towards renewable energy sources~\citep{ipcc6_wg1_synth}. 
Although these are more sustainable, they are dependent on the ever changing weather and its associated uncertainties, such as the unknowns of climate variability \citep{gernaat2021climate,Bloomfield2021nextgen,Stoop2024phd}. 

Energy system planners and operators around the world guide this transition towards renewables. 
For their work, they need insightful knowledge on the weather-driven threats that might pose a risk to the adequacy of the electricity grid in the future~\citep{gernaat2021climate,craig2022disconnect}.
A critical situation in a power system where load cannot be met by demand may not always be due to an outage of a power plant. 
It can also be due to structural shortage of energy due to the specific weather conditions at that time. 
Shortages can be related to the synoptic-scale state of the atmosphere ~\citep{vanderwiel2019regimes,Bloomfield2019regimes,vanDuinen2024covariability,wuijts2023linkingunservedenergyweather,vanDuinen2025}.
For instance, weather driven shortages can be related to dunkelflautes~\citep{vanderwiel2019extreme,Kittel2024,Biewald2025}. 
A dunkelflaute is a period of little sun and little wind. 
Insightful knowledge could be gained by better understanding the synoptic-scale atmospheric circulation patterns, their natural variability and the impact of climate change on them. 
One tool for analysing these patterns is the framework of Weather Regimes (WR). 

Weather Regimes are classifications of the meteorological variability into quasi-stationary, persistent and recurrent large-scale circulation patterns of the atmosphere \citep{michelangeli1995weather}. 
Here we will focus on WRs in the Euro-Atlantic sector.
Because the weather in the winter period in Europe is more persistent, European Weather Regimes are often defined for the (extended) winter period~\citep{michelangeli1995weather,neal2016flexible,falkena2020regimes}, although year-round definitions exist~\citep{Grams2017weather}.

The framework of WR's is a well-established technique within meteorology. 
Its roots dating back to the 1940's, when the German weather service developed a classification system for synoptic circulation patterns over Europe~\citep{dwdWetterKlima}.
These were used to understand the synoptic variability of the atmosphere, so they are not necessarily persistent patterns.
Then, since the 1990's several methods to cluster large scale patterns have been used to study circulation patterns and their impacts ~\citep{Vautard1990, Mo1987, Molteni1990}.
Discussions about the ideal number of regimes have been extensive, as this has to be set a priori~\citep{falkena2020regimes, michelangeli1995weather, Christiansen2007}. 
However, the same treatment has not been extended to other settings within the WR classification methods, such as the resolution of the grid, the period of the data and the investigated region. 

Weather Regimes are now also used to investigate the impact of meteorology within the energy sector. 
These circulation patterns over Europe influence the generation of renewables and the energy demand in Europe, through their relation to wind and temperature anomalies. 
There are indicators of a direct relation between certain Weather Regimes and periods of low potential renewable energy generation with high demand.
Especially a Scandinavian blocking pattern has been linked frequently to winter-time energy shortages in central-Europe~\citep{vanderwiel2019regimes,Bloomfield2019regimes,wuijts2023linkingunservedenergyweather}. 
Furthermore, we need to understand the impact of climate change that are relevant for future energy generation, especially because society moving towards the use of renewables. 
WR's can be used as a tool for this.~\citep{van2012vulnerability,craig2022disconnect,McKenna2022,IEA2023_variability}.  
While \textcite{Dorrington2022jetstream} show that the Atlantic jet-stream indicates that there is a trend towards less persistent blocking regimes in Europe under climate change, these results uses hybrid geopotential-jet regimes and this definition is rarely used in energy system impact assessments. 

While the framework of WR's has now taken root within the energy sector, the method often used for this is not challenged.
WR definitions are not often compared to each other. 
The classic definition~\citep{michelangeli1995weather} is used without challenge while for applications, like the energy sector where they are used to steer large public investments, it is useful to understand uncertainties of each method. 
To properly integrate climate projections into energy system applications of WR's, understanding the energy-meteorological uncertainty is required~\citep{craig2022disconnect,vanderMost2022framework,Stoop2024phd}. 
In these applications the uncertainty of the impact model itself is often overlooked.
In our case, the propagated uncertainties caused by how the Weather Regimes are defined, when they are implemented into impact models~\citep{Shepherd2019}.

In this paper we investigate the stability of WR definitions. 
Specifically to support the usability of a WR framework in energy sector impact assessments. 
For this we analyse the gaps in the knowledge of this framework by comparing the stability of three different Weather Regime definitions.
We use ERA5 reanalysis data to identify Weather Regimes, which can provide us with an understanding of the inherent uncertainty of Weather Regime definitions. 
Furthermore, the suitability of WR-methods for climate change research is scrutinized by using data from the Coupled Model Intercomparison Project in phase 6 (CMIP6) multi-model ensemble.

This paper is structured as follows. 
In Section~\ref{sec:method}, we explain how we validated our usage of the established WR definitions, how the sensitivity of the definitions was tested and how the CMIP6 models were used for the suitability analysis. 
Then, in Section~\ref{sec:results} we present the primary findings of our validation analysis, the sensitivity assessment of the WR definitions on ERA5 data and the application of the WR definitions on CMIP6 model data.
In Section~\ref{sec:discussion} we discuss the merit of our work in the larger context. 
In Section ~\ref{sec:conclusion} we present our conclusions on what these results mean for energy impact models.


\section{Method}
\label{sec:method}
As mentioned in the introduction, we need to understand the energy-meteorological uncertainties involved to determine the suitability of WR definitions for energy sector impact assessments\citep{craig2022disconnect,Stoop2024phd}.
Therefore, we need to understand how WR's are identified, why certain methodological choices are made within WR definitions, and how stable the regimes are with respect to these methodological choices.
To support this assessment the methodology is split into four sections. 

First, in Section~\ref{method:WRframework} we explain how WR's are characterized.
Second, we introduce three different WR definitions, how we apply these definitions and what their main differences are in Section~\ref{method:WRdef}.
Third, in Section~\ref{method:Sensitivity} we discuss how we tested the sensitivity of the results to changes in methodology and varying parameters. 
Finally, we explain how the suitability of the WR framework is evaluated as an impact model for the energy sector in Section~\ref{method:impactmodel}.

\subsection{Framework of Weather Regimes}
\label{method:WRframework}
The framework of WR's is based on identifying reoccurring persistent patterns that describe the low-frequency variability in highly variable (sub-)daily atmospheric data and assigning each point in the time series to one of these patterns. 
With both the identification and the assignment of the WR's methodological choices have to be made that influence the identified patterns and the regime dynamics. 
In this research we use three different methods of WR identification to study their accuracy and sensitivity (Section~\ref{method:WRdef}) applied to the ERA5 reanalysis dataset (SI Section \ref{method:ERA5}).

The most common method for finding these reoccurring patterns consists of three steps.
First, the Empirical Orthogonal Functions (EOFs) and corresponding time-series (principal components, PCs) of the wintertime $500 hPa$ geopotential height (gph) anomaly data are computed. 
Second, a $k$-means clustering algorithm~\citep{michelangeli1995weather,Jain2010} is applied to the first N PCs to separate the dataset in subsets, clusters, that are similar within, but different between them. Both the number of clusters $k$ and number of PCs have to be set beforehand. The more PCs are included, the more variance of the original dataset is retained.
In the clustering procedure, each data point in the time series is assigned to a cluster based on the distance between the cluster centre and the data. 
This is what we call 'hard' regime assignment. Occurrence then is calculated as the fraction of the number of days in each regime over the total number of days, whereas persistence is computed by counting the number of successive days in a regime and averaging those over the timeseries.

There are several alternative techniques for the assignment of data to the clusters, which are done after the clustering.
For example, one could assign a data point to a certain WR if its WR-index, as described in~\textcite{Michel2011wrindex}, is largest and has been that for at least an x number of days. 
Another technique for regime assignment is Bayesian assignment, as described in ~\citep{FalkenaBayes2023}. 
Here, the data is not assigned to one cluster or regime, but the probability that the data point belongs to a certain regime is calculated (for more details see Figure~1 of~\textcite{FalkenaBayes2023}).
In this case, computing the occurrence is done by adding the regime probabilities for each regime over the whole time period and dividing by the length of the time series.

\subsection{Weather Regime Definitions}
\label{method:WRdef}

To analyse the effect of the methodological choice, three definitions of Weather Regimes are investigated in this study.
In the following we discuss each method, their most notable characteristics, and which WR's are found when using these definitions. 
Further details and differences between our implementation and the original works can be found in SI Section~\ref{ap:settings}. 

\begin{figure}[!ph]
    \centering
    \includegraphics[width=0.8\linewidth]{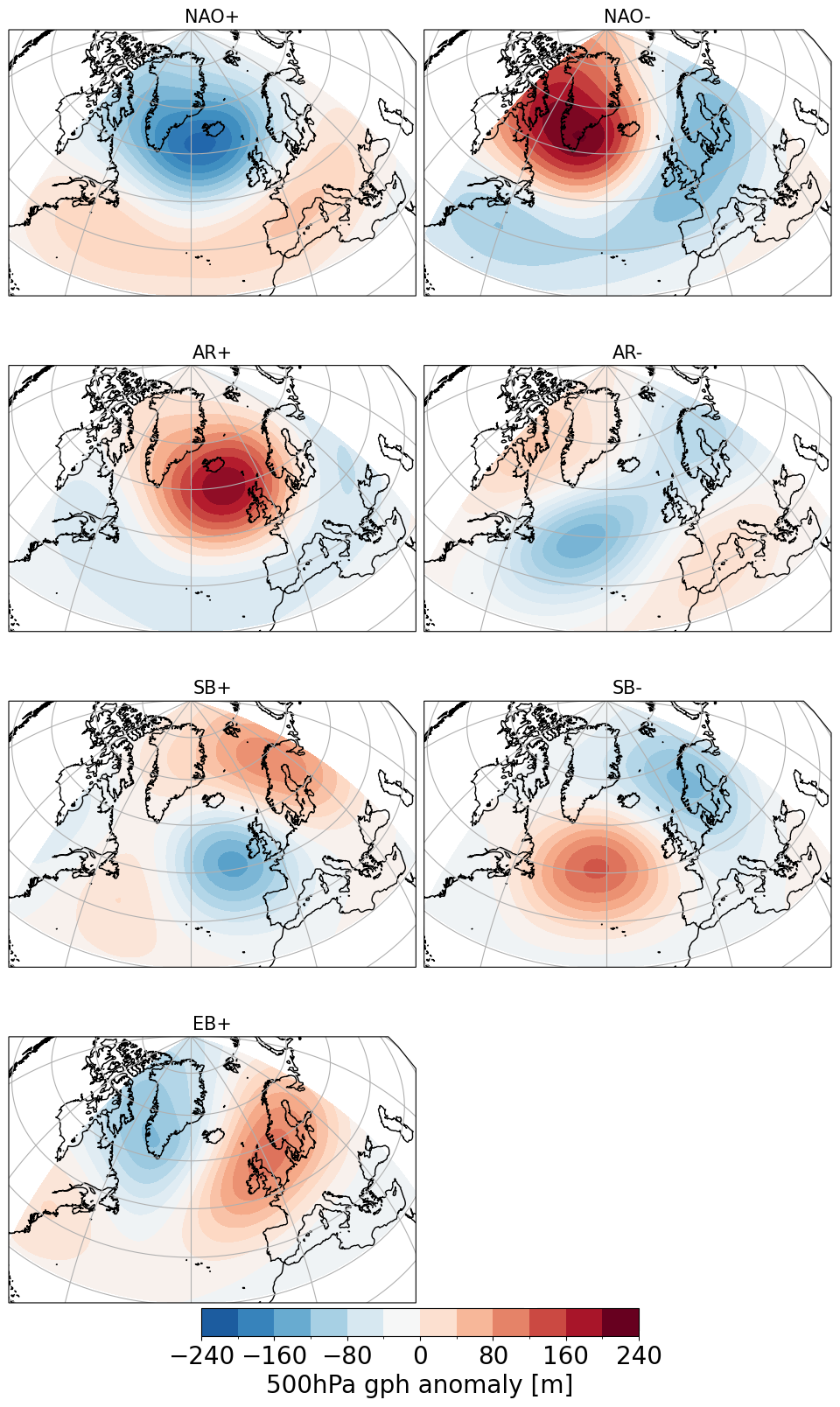}
    \caption{The 500 hPa geopotential height (gph) fields of the seven possible Weather Regimes found by replicating the method of \textcite{Grams2017weather}.
    Here blue represents negative gph anomalies and red indicates positive gph anomalies.
    The regimes here, from left to right, top to bottom are: 
    North Atlantic Oscillation positive (NAO+), North Atlantic Oscillation negative (NAO-), Atlantic Ridge positive (AR+), Atlantic Ridge negative (AR-), Scandinavian Blocking positive (SB+), Scandinavian Blocking negative (SB-) and European Blocking positive (EB+).}
    \label{fig:WRvisual}
\end{figure}

The three methodologically different methods we consider are the classic 4-regime definition \citep{michelangeli1995weather}, Falkena's 6-regime definition \citep{falkena2020regimes}, and Grams' 7-regime definition \citep{Grams2017weather}.
Although each definition consider a different number of regimes, visually similar regime patterns are found across these definitions. 
Figure \ref{fig:WRvisual} shows the seven regime patterns that are found for the WR7 definition; 
the North Atlantic Oscillation in positive and negative phase (NAO+/-), the Atlantic Ridge in positive and negative phase (AR+/-), Scandinavian Blocking in positive and negative phase (SB+/-) and European Blocking (EB+).
The regimes identified by the WR4 definition are NAO+, NAO-, AR+ and SB+. For WR6 these four regimes are found together with AR- and SB-.

The spatial patterns of the regime patterns do differ slightly between the different WR definitions (see Figure~\ref{fig:WR4base} and Figure~\ref{fig:WR6base}). 
It is important to note that the names describing the WRs vary in literature, where here we standardised the naming based on the terminology used in~\textcite{falkena2020regimes}.
In Table \ref{tab:WRnames}, the equivalent names that are used in \textcite{Grams2017weather} can be found.

The classic wintertime 4-Regime (WR4) definition is based on \textcite{michelangeli1995weather}. It is the most commonly used WR definition, so literature discussing this definition is extensive~\citep[e.g.][]{Straus2007,Dawson2012,Hannachi2017} and the connection between these regimes and the energy sector has been well documented ~\citep[e.g.][]{vanderwiel2019regimes,Bloomfield2019regimes,GarridoPerez2020}. 
However, due to the widespread application of the four regimes, not one universal methodology has been established. 
Many factors between the identification methods used in literature differ, which can have an effect on the outcome (most notably the persistence). 
Here, the method described in \textcite{vanderwiel2019regimes} is followed.
They use the first 14 PCs of the DJF Z500 data to identify the four WRs using k-means clustering.

For Falkena's wintertime (DJFM) 6-regime definition (WR6), we use the method described in their initial work on weather regimes; \textcite{falkena2020regimes}.
This definition stands out as no EOFs are used, eliminating the practice of double filtering of the data and finding 6 regimes to be optimal instead of 4.
Within the WR6 method the grid-point data is directly clustered into 6 clusters with a $k$-means clustering algorithm. 
Since this is a new method, its use is less established within energy-meteorology. 

For Grams' 7-regime definition (WR7) we aimed to replicate the method described in \textcite{Grams2017weather}. 
Where the previous two methods focus on the winter months, this method analyses the recurrent patterns throughout the year and is applicable year-round.  
This last aspect is ideal for energy sector applications and this methodology has been used in various energy sector impact assessments~\citep{Grams2017weather,Mockert2023regimelowwind}.
In this method, first the Z500 anomaly data is obtained by subtracting the monthly mean (eliminating seasonal variations) after which the leading 7 PCs are clustered into 7 clusters.
A significant methodological difference between this method and the other WR definitions, is that a 10-day low-pass filter was used in the original work. 
However, we found this to be too computationally heavy to use within the WR framework assessment presented here. 
Furthermore, this definition uses a `No regime' designation, which we also have not been able to replicate. 

For each of the three WR definitions, details on the implementation used here, differences with the original method and further literature are discussed in SI Section \ref{ap:settings}.
We validated the WRs for each definition by visually comparing the $500 hPa$ gph anomaly fields of the clusters we found to the gph anomaly fields in the relevant literature for each WR definition.
When the occurrence and persistence of the WR's was given in the original work, we also compare those to our findings.

\subsection{Sensitivity assessment of WR framework}
\label{method:Sensitivity}
\begin{figure}[!t]
    \centering
    \includegraphics[width=
    \linewidth]{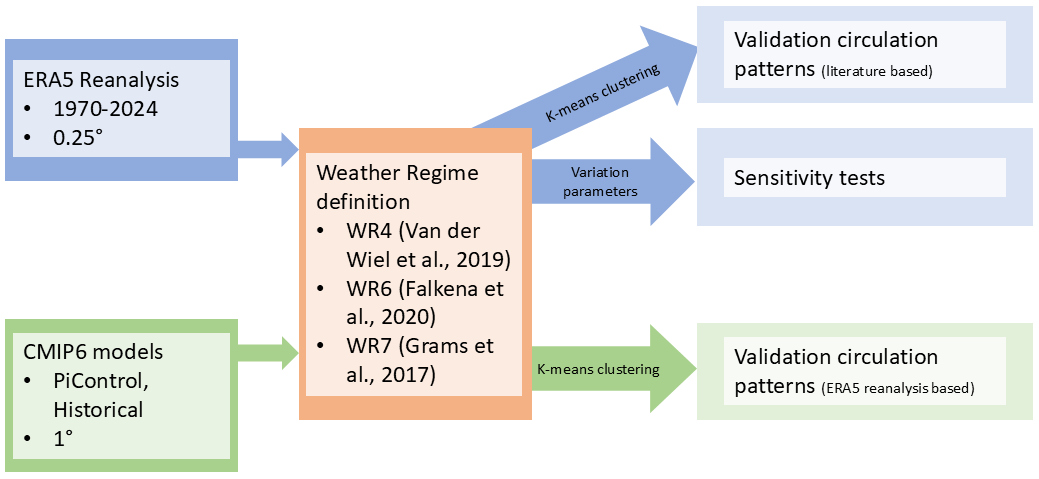}
    \caption{General overview of the data processing chain of determining the Weather Regimes (WR) and their occurrence and persistence for ERA5 reanalysis and CMIP6 model data. 
    The number after `WR' indicates the number of regimes found and the reference indicates on which literature the method is based. 
    The blue arrows indicates how the ERA5 reanalysis data is used. 
    The green pathway shows how the CMIP6 model data is handled.
    The orange box represents the process of calculating the regimes, that is identical for both data sources.}
    \label{fig:dataprocess}
\end{figure}

The aim of this study is to analyse the gaps in knowledge, and the suitability, of the WR framework for energy sector impact assessments. 
For that, understanding the sensitivity to parameters inherent to the WR definition and data is vital. 
To assess this we study the effect of varying two types of parameters, WR definition and data parameters, on the occurrence, average persistence and stability of the regimes.
Here, stability refers to the ability to identify the expected regimes. To verify this we use automatic matching of the regimes of each sensitivity experiment to those of the baseline. The process of this is explained in more detail in SI Section \ref{ap:regimesautomatic}. 
The steps of the data processing chain for the sensitivity analysis are visualised in the blue stream of Figure~\ref{fig:dataprocess}. 


WR definition parameters that we have varied can be divided into two categories:
First the number of EOFs.
For WR4 and WR7 the number of EOFs used in the first step of calculating the regimes was varied between 7, 14 and 30 EOFs.
This was done because using more EOFs means more of the variance of the dataset is being considered. 
So, understanding how the number of EOFs influences the persistence or occurrence of certain regimes, can inform which number would be best practice. 

Second, the effect of changing the method of assignment, hard or Bayesian, is investigated.
The assignment of each data point to the closest cluster is a straightforward technique, but it is not a realistic representation of the atmospheric circulation.
Therefore, we also looked into the effect of using a probabilistic method of assignment. 

Understanding how the specific data the regimes are clustered on affects the occurrence and persistence is necessary, because weather is extremely dynamic.
If the results are highly dependent on the data, this implies instability within the method. 
Especially because the data-sources vary across methods.
Data parameters that we have varied can be divided into two categories;
time dependent parameters and space dependent parameters.

The time dependent parameters we consider are the time period, time resolution and which months are taken into account. 
We look at the time period to understand the stability of the regime definitions across decades.
This is done with two different approaches:
The first approach is to both cluster and assign the data on the same time period. 
Another approach we use is to cluster the data on the data from 1971--2020, the full available period, and then assign the clusters to data in the 30-year climatological periods 1971--2000, 1981--2010 and 1991--2020. 
We also research the effects of altering the time resolution, 6-hourly or daily data, to assess whether or not more data translates into a more accurate representation of reality. 
Furthermore, which winter months are used varies between the WR4 and WR6 definitions.
So we investigate if there is a notable difference to persistence or occurrence of certain regimes if in addition to December, January and February being considered, March is included. 
During this process, the anomalies are calculated with the same method; for each day the average for that specific day is subtracted.

Space parameters that are investigated are the grid resolution and the region. 
Because grid resolution greatly affects computational costs, the effect of changing the grid resolution from $1^\circ$ to $2.5^\circ$ can vastly reduce these costs. 
Therefore, understanding if reducing the grid resolution has an influence on the quality of the results is important. 
Various definitions define the Euro-Atlantic region differently.
So, we need to understand how defining the region differently affects the geographic location of low- and high-pressure areas within the regimes and how this affects the persistence and occurrence of the regimes.
Therefore we vary the region of the data the regimes are clustered on between the region used in ~\citep{vanderwiel2019regimes}
(90$^\circ$W-30$^\circ$E, 20$^\circ$-80$^\circ$N), the region used in ~\citep{Grams2017weather} (80$^\circ$W-40$^\circ$E, 30$^\circ$-90$^\circ$N), and three regions stated in SI Section \ref{ap:sensdataparam}, to investigate the influence of different Western, Eastern and Northern bounds.

\subsection{Sensitivity of WR framework within climate models}
\label{method:impactmodel}
To assess the suitability of the WR framework for applications to impact models in the energy sector, we need to assess the stability of the framework when applied to climate model data that include projections of future scenarios.
Therefore, we apply the framework of WR's to climate model data to evaluate its applicability for energy sector impact assessments. 
To assess the impact of climate change, the aleatoric, epistemic, scenario, and as \citet{Shepherd2019} described, the impact model uncertainty, needs to be assessed in the same manner. 
The aleatoric, or year-to-year, uncertainty is addressed by the variation of the time periods used (Section \ref{method:Sensitivity}) and the uncertainty of the impact model by using three methodologically different WR definitions. 
So, to facilitate climate change impact assessments we need to address the emission scenario and the epistemic (e.g. the climate model) uncertainty. 

Epistemic uncertainty is tested by applying each of the three WR definitions a sub-ensemble of the Coupled Model Intercomparison Project phase 6 (CMIP6).
This process is visually represented in Figure \ref{fig:dataprocess} by the green arrows. 
Our sub-ensemble consisted of seven climate models chosen due their data availability covering the set of the simulations needed, an overview of their references is shown in Table \ref{tab:cmip6models}.
We refer to each combination of CMIP6 model and either piControl or historical data, as one CMIP6 experiment. 
To assess suitability of our sub-ensemble of CMIP6 models we applied the framework of WR's and calculating the distance between the regimes found in a CMIP6 experiment and the regimes found in the ERA5 reanalysis data. 
For each of the WR definitions the standard settings, also used for the sensitivity tests, are used as provided in SI Section~\ref{ap:settingsvalera5}, with a few exceptions.
These exceptions are shown in Table \ref{tab:CMIP6ERA5baseline} and we call these the 'baseline' settings. 
The regimes from the reanalysis data that we will use to compare are also found by using the baseline settings, which we then call the baseline regimes.
Then the regimes from the CMIP6 experiment are matched to the ERA5 regimes using the process described in SI Section \ref{ap:regimesautomatic}.
When evaluating if a WR definition produced the expected regimes when clustered on CMIP6 model data, we consider it successful if the normalised distance between the baseline regimes and the CMIP6 experiment regimes is $<0.4$. 
This value is chosen because we want the WR definition to perform better than an empty array, which can produce the value of 0.475.

Emission scenario uncertainty can be tested by assessing the differences in occurrence and persistence of the regimes in different Shared Socio-Economic Pathways (SSP) data streams from the CMIP6 models. 
Establishing a difference in persistence and occurrence of regimes between historical and SSP126, SSP370 and SSP585 data streams could be an indication of a change due to climate change. 
Which in turn could inform impact models used in the energy sector. 
So, for our CMIP6 models ensemble the persistence and occurrence of each of the regimes found by all three WR definitions using Bayesian assignment are calculated. 
However, due to difficulties with reproducing the expected regimes, these results will not be discussed in this paper.


\section{Results}
\label{sec:results}
Here we discuss our analysis of the weather regime framework. 
First, we discuss the results of our validation of the methods described in Section~\ref{method:WRdef}.
Then we delve into our findings of the sensitivity of these methods to various assumptions  for each of the definitions used. 
Finally, we discuss the results of the sensitivity of the WR framework within the context of climate change, where we trained the WR models on CMIP6 model data.

\subsection{Validation of Weather Regime implementation}
\label{res:validation}
We performed a validation of our WR implementation by comparing the results produced by our WR definitions to the results of the corresponding original method.
For each WR definition we will discuss how they perform in terms of reproducing the regimes, and where possible, compare the occurrence and persistence rates of the regimes to the corresponding literature values.
For this the models were set with parameters as close to the original method as possible. 
It is not possible to perform statistical analysis on how close the regimes are to the original method, for any of the definitions, because the gph anomaly maps of the original regimes were not available to us.

The regimes found with our WR4 model closely resemble the regimes found by \textcite{vanderwiel2019regimes}. 
Figure \ref{fig:WR4base} in the SI shows the four regimes found by using the WR4 model with the parameters shown in SI Table \ref{tab:WR4settings}. 
Visually, the patterns are similar.
Small differences in geography and intensity of the highs and lows can be identified, but no large discrepancies are present.
Furthermore we found that, occurrence rates of weather regimes for the WR4 regimes are at most $1\%$  different from the rates found by ~\textcite{vanderwiel2019regimes}.
This can be seen in Table \ref{tab:validationWR4} in SI Section \ref{ap:validoccpers}.
Persistence rates were not available in~\textcite{vanderwiel2019regimes}, so these cannot be compared. 

The regimes found with our WR6 model do resemble the regimes found by \textcite{falkena2020regimes}. 
Figure \ref{fig:WR6base} in the SI shows the four regimes found by using the WR4 model with the parameters shown in SI Table \ref{tab:WR6settings}. 
Visually, the patterns of the highs and lows are located over the same regions.

SI Table \ref{tab:validationWR6} in Section \ref{ap:validoccpers} shows that our persistence rates for the WR6 definition are at most 0.6 days lower than the persistence rates found in \textcite{falkena2020regimes}. 
Furthermore, this Table shows that the difference of occurrence rate of the regimes we calculate and the regimes from ~\textcite{falkena2020regimes} are within $2.2 \%$. 

Visually comparing the regimes found here to the regimes produced in \textcite{Grams2017weather}, we see that the patterns of the highs and lows are in similar geographical positions.  
Attempts to recreate a `no regime' designation failed, or produced vastly different rates of occurrence compared to \textcite{Grams2017weather}.
So we decided to not include the `no regime'.

This lack of a `no regimes' means a comparison of occurrence rates is not possible for the other regimes, as this rate is influenced by the occurrence of the `no regime'. 
Persistence rates were not given in \textcite{Grams2017weather}. 
The persistence and occurrence rates we found, are shown in SI Table \ref{tab:validationWR7}.

\subsection{Sensitivity of WR definitions}
\label{res:sensitivity}
Within the energy sector the scale of the large public investments require a thorough understanding of the uncertainties in methods that steer them. 
Therefore we performed an analysis of the sensitivity of the WR framework to both methodological and data selection choices. 
As a result of our deep-dive into the original implementations and our tuning of the implementation within validation process we developed a robust methodology of three WR definitions, on which we have performed sensitivity tests. 
The complete analysis of sensitivity with additional findings and figures can be found in SI Section~\ref{ap:figuressens}. 

The effect of altering the variables mentioned in Section \ref{method:Sensitivity}, on the occurrence and persistence of Weather Regimes was tested.  
First we will discuss the results of the parameters that alter the WR definition itself: the number of EOFs and the method of assignment. 
Then, the effect of altering the input data by varying data parameters is investigated. 
There we will focus on the effect changing the period the clustering algorithm is trained on, because this causes the largest instabilities and uncertainties. 

Figures that include the impact of the WR definition dependent parameters are included in SI Section \ref{ap:senswrdefparam} for each WR definition. 

Figure \ref{fig:WR4sensmodel} shows that changing the number of EOFs does not have a large effect for the WR4 definition. 
However, for the WR7 definition, Figure \ref{fig:WR7sensmodel} indicates that a larger number of EOFs increases the persistence for some regimes.
The difference between 14 and 30 EOFs is small, so the largest difference is between 7 and 14 EOFs. 
This implies that the leading 7 EOFs do not capture all the relevant variance.
\begin{figure}[!t]
    \centering
    \includegraphics[width=1\linewidth]{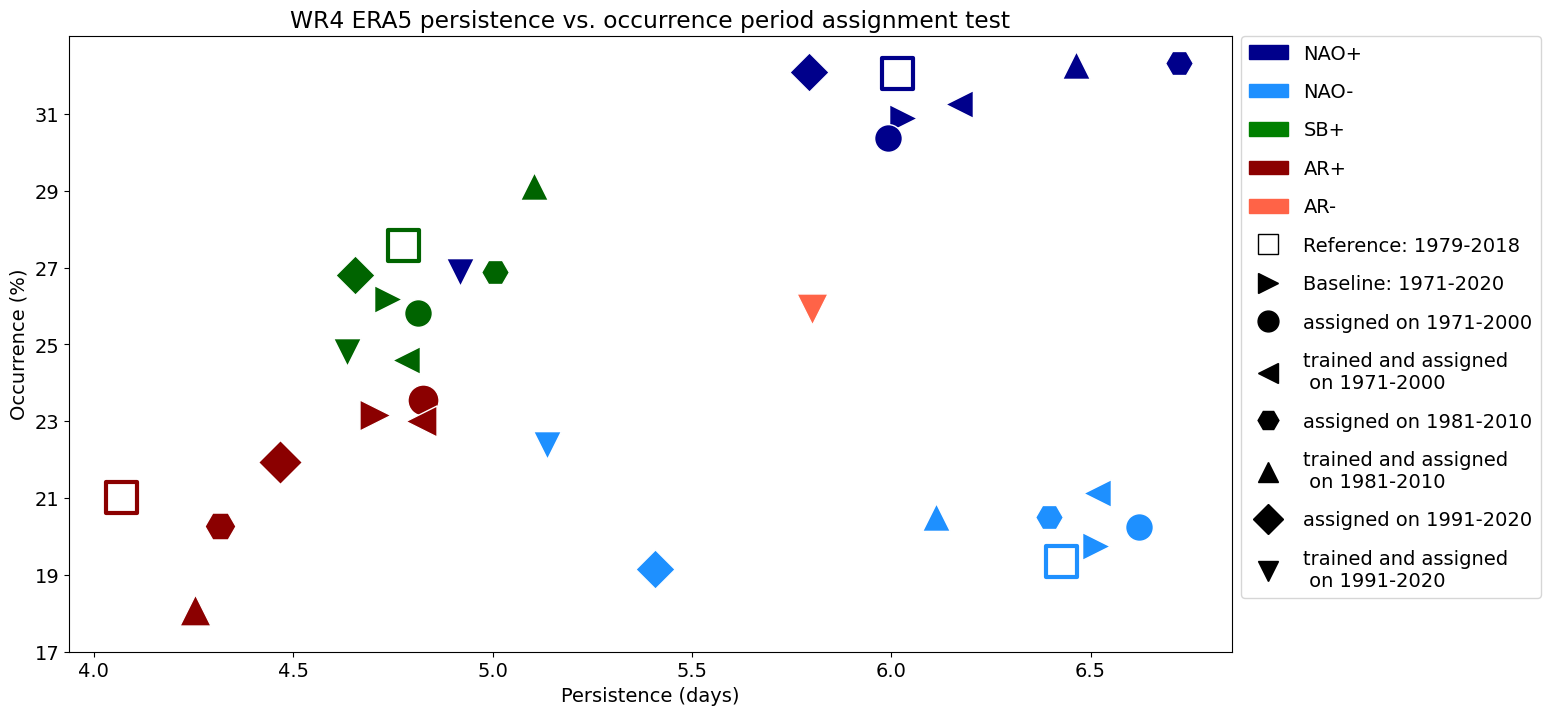}
    \caption{Persistence and occurrence for the identified Weather Regimes (WR) using the  WR4 definition are represented by colour.
    The North Atlantic Oscillation positive (NAO+) is represented by dark blue and the North Atlantic Oscillation negative (NAO-) is represented by light blue.
    Scandinavian Blocking positive (SB+) is represented by green.
    Atlantic Ridge positive (AR+) is dark red and Atlantic Rdige negative (AR-) is orange. 
    The method is carried out on varying time periods (shape), where for certain experiments the regimes are identified within the baseline period, 1971--2020, and assigned on a different period, or identified and assigned on the same period. Here the reference period is the period used in \textcite{vanderwiel2019regimes}.
    Note that for only one time period, AR- instead of its positive phase, is found.
    }
    \label{fig:WR4periods}
\end{figure}

Changing the method of assignment from hard assignment to Bayesian assignment increases the persistence for all regimes for each WR definition. 
This change affects the occurrence rate per regime differently for each WR definition. 
For WR4 only a small change in occurrence can be identified, as seen in Figure \ref{fig:WR4sensmodel}. 
Figure \ref{fig:WR6sensmodel} shows that for WR6, an increased occurrence for NAO+ and NAO- and a decreased occurrence for SB+ and AR+ can be identified. 
WR7 also shows an increased occurrence for NAO+ and NAO-, but a decreased occurrence for SB+, SB- and AR-, which is shown in Figure \ref{fig:WR7sensmodel}.
Note that when calculating the persistence of the regimes while using Bayesian assignment, the WR with the highest probability is assigned to each data point.

Figures that include the impact of the data parameters, are included in SI Section \ref{ap:sensdataparam} for each WR definition: 
Figures \ref{fig:WR4sensdata}, \ref{fig:WR6sensdata} and \ref{fig:WR7sensdata} for WR4, WR6 and WR7, respectively. 
Those figures show that grid resolution does not have a large effect on either persistence or occurrence.

Furthermore, the temporal resolution only affects the persistence, which is to be expected when using hard assignment. 
More data points means a larger chance to switch to a different regime faster. 
So, for all WR definitions and all regimes the persistence is decreased when switching from daily to 6-hourly data. 

In addition to that, changing the winter months from December--February to include March, has a small, but noticeable effect for some regimes. 
For the WR4 definition, NAO- has a higher occurrence and longer persistence when March is included and SB+ has a lower occurrence, as can be seen in Figure \ref{fig:WR4sensdata}. 
However, Figure \ref{fig:WR6sensdata} shows that for the WR6 definition, the difference for NAO- is negligible when March is excluded.
Here the regimes SB+, SB- and AR- have the most notable shift in occurrence and/or persistence. 

Moreover, the extent of the region does affect the persistence and occurrence of each of the regimes for all WR definitions.
This is explained further in SI Section \ref{ap:sensdataparam} and can be seen in Figures \ref{fig:WR4sensdata}, \ref{fig:WR6sensdata} and \ref{fig:WR7sensdata}, for WR4, WR6 and WR7, respectively. 

\begin{figure}[!t]
    \centering
    \includegraphics[width=0.9\linewidth]{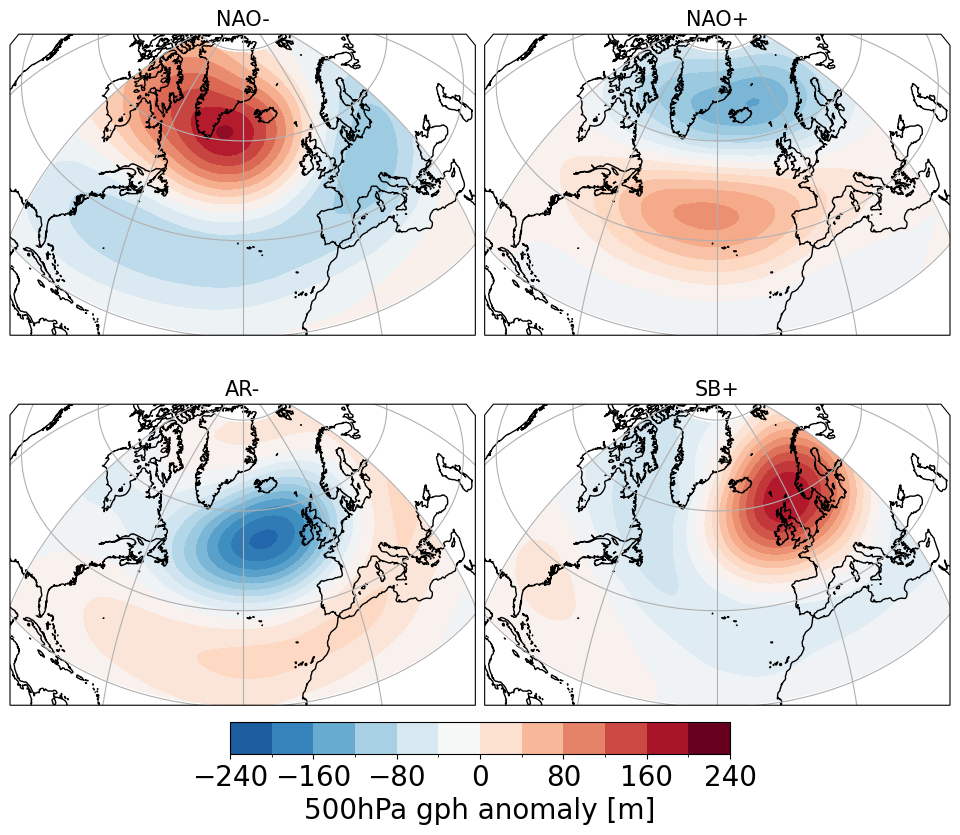}
    \caption{500 hPa geopotential height (gph) anomaly fields of the four Weather Regimes (WR) found by applying the WR4 definition, based on \textcite{vanderwiel2019regimes}, by training the clustering algorithm on the period 1991-2020.
    Here, red indicates a positive gph anomaly and blue indicates a negative gph anomaly.
    Instead of finding the expected Atlantic Ridge positive regime, its negative phase is found.
    }
    \label{fig:WR41991-2020pattern}
\end{figure}

The largest effects can be found in altering the period.
So, we will now focus on how the models are affected by changing the period of the data the models cluster the regimes on. 
Figure \ref{fig:WR4periods} shows the occurrence and persistence of the WR4 definition trained on ERA5 data, where the clustering algorithm is trained on different time periods. 
The reference period, 1979--2018, is the period used in both ~\citep{vanderwiel2019regimes} and ~\citep{falkena2020regimes}.
We call 1971--2020 the baseline period.
This period was chosen, because it is a longer, presumably more stable, climatological period.
A distinction is made between training the clustering algorithm on a specific period and assigning those clusters to that period, or training the clustering algorithm on the period 1971--2020 and only assigning the clusters on a specific period. 
There is a large spread for all four regimes in both occurrence and persistence, depending on training and/or assigning the regimes on different time periods. 
Special attention needs to be given to the result of training the clustering algorithm on data of the period 1991--2020. 
A visualization of the patterns found here is shown in Figure \ref{fig:WR41991-2020pattern}. 
There we can see that no Atlantic Ridge regime can be identified.
Instead, a regime resembling its negative phase is found. 
Linked to this change, the NAO+ is notably different from previously found patterns.

In addition to the instability in time for the WR4 definition, both the WR6 and WR7 definitions are strongly affected by the period in which the clustering algorithm is trained. 
Even though all expected WR's can be identified, for the WR6 definition, a large spread in occurrence and/or persistence is found, depending on the period considered.  
This can be seen in Figure \ref{fig:WR6persoccperiod} in SI Section~\ref{ap:sensdataparam}. 

Furthermore, the maximum anomalies of the WR7 Weather Regimes shift visibly, based on different time period the regimes were clustered on. 
Figure \ref{fig:WR71991-2020pattern} shows the patterns of the regimes that are found by clustering the data from the period 1991--2020. 
There it is can be seen that the found regimes are different from the reference regimes. 
This is especially the case for the AR- regime, where the maximum gph anomaly has shifted from the Atlantic ocean to lying over Great Britain. 
Due to a lack of computational power, it was not possible to train the model on the period 1971--2020 and assign the regimes to different time periods for the WR7 definition.

\subsection{Sensitivity of WR framework within climate models}
\label{results:CMIP6}
To assess the suitability of the WR framework for applications to impact models in the energy sector, we need to assess the stability of the framework when applied to climate model data that include projections of future scenarios.
With stability we mean that the WR definitions reliably produce the expected regimes for different types of input data. 
As described in Section~\ref{method:impactmodel}, we used a sub-ensemble of CMIP6, and standard parameters within the WR framework. 

Instability of the WR definitions is found when applying the three definitions on various CMIP6 models. 
Whether or not the expected Weather Regimes are found depends on which model and data set (piControl or historical) is used. 
Table \ref{tab:CMIP6results} shows for which combinations of CMIP6 models, data set and WR definition, the expected Weather Regimes can be identified.  
For normalised distances $<0.4$ we consider the model to perform well enough to produce the expected regimes and the cells are coloured green. 
When the normalised distance is $>0.4$ the distances to the respective ERA5 reanalysis baseline regimes are too large to be considered the same regimes, so there the model/data set combination is coloured red. 
Blue cells show data streams that were empty. 
Furthermore, white cells indicate that no successful clustering has been performed yet and/or the distance between the ERA5 reanalysis data and the model output has not been calculated.\footnote{In the case of MPI-ESM1-2-LR piControl, the data set was empty, so no distance was calculated.}

WR4 seems to be relatively stable when applied to the CMIP6 model data.
Of the twelve CMIP6 experiments, ten produce the expected regimes. 
For WR6, nine CMIP6 experiments were executed, where none performed well enough to stay below the threshold of the normalised distance from the ERA5 reanalysis regimes.  
Of the nine CMIP6 experiments performed with the WR7 definition, all of them have a normalised distance to the ERA5 reanalysis regimes smaller than 0.4. 

However, for certain experiments, when visually comparing the regimes to ERA5 reanalysis, not all of the regimes found in the data stream can be confidently matched to an ERA5 regime. 
For instance, when applying the WR7 definition to the NorESM2-MM Picontrol data, the normalised distance is 0.349. 
Despite the distance passing the threshold, visual inspection shows substantial differences. 
The regimes produces by this experiment are shown in SI Figure \ref{fig:NorESM2piControlWR7}, where there is no clear AR+, but instead a regime resembling SB+ is found. 
Additionally, the EB+ regime is located further towards the Atlantic Ocean than in Figure \ref{fig:WRvisual}.
Similar behaviour, the shifting of regimes or not finding a regime clearly, is present in three other experiments for the WR7 definition. 
These experiments are: NorESM2-MM historical, MPI-ESM1-2-LR historical and UKESM1-0-LL historical.

\begin{table}[!t]
\centering
\caption{The three Weather Regime (WR) definitions (WR4, WR6 \& WR7) applied on piControl (piC) and historical (hist) data streams of seven CMIP6 models. 
Here it is assessed whether or not the expected regimes are found when the clustering algorithm is trained on the indicated data stream. 
For normalised distances $<0.4$ we consider the model to perform well enough to produce the expected regimes and the cells are coloured \textcolor[HTML]{32CB00}{green}. 
When the normalised distance is $>0.4$ the distances to the respective ERA5 reanalysis regimes are too large to be considered the same regimes, so there the model/data stream combination is coloured \textcolor[HTML]{FD6864}{red}. 
The \textcolor[HTML]{00D2CB}{blue} cells indicate that for that model data stream the dataset was empty. 
Empty cells for the CESM2 data streams indicate that no successful clustering has been performed yet.
For the UKESM1-0-LL and MPI-ESM1-2-LR no distance could be calculated in a reasonable time frame, but the clustering was performed with varying success.}
\label{tab:CMIP6results}
\begin{tabular}{|l|l|l|l|l|l|l|}
\hline
                 & \rot{WR4 piC}                                              & \rot{WR4 hist}                                             & \cellcolor[HTML]{FFFFFF}\rot{WR6 piC} & \rot{WR6 hist}                      & \rot{WR7 piC}                       & \rot{WR7 hist}                      \\ \hline
CMCC-CM2-SR5     & \cellcolor[HTML]{FD6864}0.430                        & \cellcolor[HTML]{32CB00}0.361                        & \cellcolor[HTML]{FD6864}0.437   & \cellcolor[HTML]{FD6864}0.585 & \cellcolor[HTML]{32CB00}0.286 & \cellcolor[HTML]{32CB00}0.319 \\ \hline
CESM2            & \cellcolor[HTML]{32CB00}{\color[HTML]{000000} 0.280} & \cellcolor[HTML]{32CB00}{\color[HTML]{000000} 0.256} &                                 &                               & \cellcolor[HTML]{FFFFFF}      &                               \\ \hline
NorESM2-MM       & \cellcolor[HTML]{32CB00}{\color[HTML]{000000} 0.214} & \cellcolor[HTML]{32CB00}{\color[HTML]{000000} 0.217} & \cellcolor[HTML]{FD6864}0.444   & \cellcolor[HTML]{FD6864}0.540 & \cellcolor[HTML]{32CB00}0.349 & \cellcolor[HTML]{32CB00}0.280 \\ \hline
NorESM2-LM       & \cellcolor[HTML]{FD6864}0.421                        & \cellcolor[HTML]{32CB00}0.276                        & \cellcolor[HTML]{FD6864}0.679   & \cellcolor[HTML]{FD6864}0.468 & \cellcolor[HTML]{32CB00}0.375 & \cellcolor[HTML]{32CB00}0.341 \\ \hline
UKESM1-0-LL      & \cellcolor[HTML]{32CB00}0.230                        & \cellcolor[HTML]{32CB00}0.372                        &                                 & \cellcolor[HTML]{FD6864}0.556 & \cellcolor[HTML]{32CB00}0.239 &                               \\ \hline
MPI-ESM1-2-LR    & \cellcolor[HTML]{00D2CB}0.552                        & \cellcolor[HTML]{32CB00}0.303                        & \cellcolor[HTML]{00D2CB}        & \cellcolor[HTML]{FD6864}0.447 & \cellcolor[HTML]{00D2CB}0.475 & \cellcolor[HTML]{32CB00}0.368 \\ \hline
EC-Earth3-Veg-LR & \cellcolor[HTML]{00D2CB}0.552                        & \cellcolor[HTML]{32CB00}0.277                        & \cellcolor[HTML]{00D2CB}0.850   & \cellcolor[HTML]{FD6864}0.426 & \cellcolor[HTML]{00D2CB}0.475 & \cellcolor[HTML]{32CB00}0.344 \\ \hline
\end{tabular}
\end{table}

An example of the clustering using the WR6 definition of a data stream of a CMIP6 model, piControl of NorESM2-LM,  is shown in figure \ref{fig:WR6CMIP6NorLM}. 
Here some regimes are the regimes we expect to see, like the NAO- and the regime labeled as NAO+ by the matching algorithm resembles SB+. 
However, the other regimes have vastly different structures compared to SI Figure \ref{fig:WR6base}.
For instance, there is no regime with an intense low or high over the central Atlantic Ocean, as is expected of the AR- and AR+ regimes. 
For most CMIP6 experiments that do not pass the normalised distance threshold, some of the expected patterns can be identified, but others are similar to other regimes that have shifted or they are completely different patterns.
For WR6 the regime that is absent most often is the SB- regime.
However, the NAO- and NAO+ are reliably found by the WR6 definition across models and data sets. 
For WR4 AR+ is missing for CMCC-CM2-SR5 piControl, but for NorESM2-LM piControl all regimes have a different pattern to the expected regime. 

There is no strong evidence that training the clustering model on either piControl data or historical data has an influence on the stability of the regimes. 
The performance of the WR4 definition in the historical stream, does suggest that the historical streams produce regimes more reliably.
However, this is not the case for the WR6 or WR7 definitions, so more climate models need to be investigated to be able to draw a conclusion.


\section{Discussion}
\label{sec:discussion}
In this study various methodological choices were made that affected the outcome of the research.
Here we will explain why certain choices were made and how these affected our results. 
Moreover, we will place our results of the sensitivity tests into the context of the larger climate system and discuss the implications for impact models used in the energy sector.
Then we discuss important issues regarding the application of the WR framework on CMIP6 data. 
Finally, this Section concludes with suggesting further steps to improve and expand this research.

\subsection{Methods}
Even though the reproduction of the WR definitions was carried out with great attention to the original methods, some differences within our method exist. 
Especially problematic was the inability to reproduce the `No regime' definition from \textcite{Grams2017weather} (WR7).
Moreover, it was not possible to produce an adequate low-pass filter for the WR7 definition.
While a basic filter was implemented, the number of re-runs our sensitivity assessment of the WR framework required and the computational burden that even this inadequate filter had, forced us to not include this filter due to time available. 

Furthermore, as discussed in SI Section \ref{ap:settings}, here the geopotential height (gph) anomalies are calculated differently for WR6 and WR7, and likely also for WR4. 
\textcite{falkena2020regimes} subtracts a background state from the $500 hPa$ gph data.
In \textcite{Grams2017weather} the anomalies are calculated with respect to a 90-day running mean as reference climatology.
For WR4 it is not stated how the anomalies of the $500$ hPa gph fields were calculated in \textcite{vanderwiel2019regimes}. 
In our approach, for each day the climatology of that day is subtracted from the $500 hPa$ gph data. 
Using the climatology as the baseline should account for the slow seasonal variability without a-priori assumptions on its properties, while providing a smooth background signal. 

Furthermore, when applying the WR7 definition on the CMIP6 data we chose to alter the considered region. 
This was done because the original method includes the $90^\circ$N as a boundary, which consists of one data point. 
This would skew the results, as one data point would have a larger relative influence, so we opted to standardise the regions across definitions. 

All other discrepancies between the original method and the method carried out here, can be found in SI Section~\ref{ap:settings}. 
The various small, and sometimes unknown, differences in the model definitions are the likely culprit for the found differences for both the occurrence and the persistence rates within our validation experiments. 

Many works do not carry out a validation step within their WR framework.
In general the incomplete descriptions, the lack of published open data, and the absence of (reproducible) code (see ~\textcite{vanderwiel2019regimes} and ~\textcite{Grams2017weather}), means that a replication of these their findings is a demanding effort.  
This makes replication difficult and is a roadblock towards standardisation across implementations. 
Therefore we encourage transparency regarding the methodology of the WR framework and the use of the FAIR guiding principles~\citep{Wilkinson2016}.
See SI Section~\ref{ap:open} for our data \& code references.

\subsection{Sensitivity of WR framework}

The results of our sensitivity tests imply a large dependency on the time period the WR method clusters on. 
So, short time periods should be avoided when implementing WR definitions, as they cause large uncertainties. 
Especially considering the variability of the climate system at various shorter timescales.

For instance, when considering the effects of the multi-year El Ni\~no Southern Oscillation (ENSO), if the considered period is too short the effects of which El Ni\~no cycles are included may unintentionally influence the results. 
As found by ~\textcite{Falkena2021}, there is a relation between occurrence rates of regimes and the El Ni\~no Southern Oscillation on interannual time-scales. 
Furthermore, multidecadal variability has been observed in German wind energy generation~\citep{Wohland2019}, and European solar energy generation~\citep{Wohland2020}. 
Therefore, we recommend using periods of at least 50 years, when training WR definitions for climate system dependent implementations, such as the energy sector. 

Large scale atmospheric structures in the Euro-Atlantic sector are strongly related to the Atlantic Meridional Overturning Circulation (AMOC).
Variations in the AMOC have been linked to changes in some regime properties~\citep{Peings2014,Bellomo2023}.
As the probability of a AMOC collapse is not insignificant~\citep{Chapman2024,Smolders2024}, a possible collapse should be accounted for investigating the effects of climate change on WR's. 
\textcite{vanWesten2025} have applied the WR6 definition to CESM simulations with or without an AMOC.
While they found little change in the overall patterns or frequency of the regimes, they also found changes in the precipitation patterns for a scenario without an AMOC. 
Furthermore, \textcite{vanWesten2025temp} showed that an AMOC collapse would be associated with significant cooling during the winter months, likely leading to more stress on the energy sector due to the sensitivity of electricity demand to temperature.
Future research on the effects of climate change on WR's should be aware of this. 

\subsection{Application WR framework on CMIP6 data}

The distances calculated in Section~\ref{results:CMIP6} to validate the regimes found in the CMIP6 data streams should be treated with caution. 
The reason for this is that for some data streams, across WR definitions, normalised distances above 1 were calculated within the distance matrix. 
We were unable to determine where this error comes from.
Thus the specific value of the distance should not be taken at face value.
However, as the same method was used for each experiment and the results are in line with what can be concluded visually we assume the calculated distances are still an indication of the quality of the found regimes.

Furthermore, there are four data streams where the regimes produced by the WR7 definition pass the 0.4 normalised distance threshold, but the regimes are notably different to the corresponding ERA5 reanalysis regimes.
This is discussed in further detail in SI Section \ref{ap:regimesCMIP6}.
When applying the WR7 definition to CMIP6 data streams, great caution should be taken when assessing the performance of the clustering. 

Keen readers might have noticed the lack of the persistence and occurrence results of the regimes found in CMIP6 Shared Socio-economic Pathways data streams.
While the data was retrieved, and the WRs identified (where possible, see Table~\ref{tab:CMIP6results}), the results would be prone to misuse.
We opted not to show these as we found that the WR framework is unsuitable for such calculations, due to the instability of the regimes found.

\subsection{Outlook}
This research could be improved and expanded in the following two manners.
First, when carrying out the sensitivity tests, a Bayesian assignment, as described in \textcite{FalkenaBayes2023}, could be used for all the tests. 
Here we only investigated the results of implementing the Bayesian assignment itself, but varying other data parameters when using a more stable assignment would alter the occurrence and persistence rates.
Assessing how sensitive the definitions are, when a more persistent method of assignment is used, could possibly reduce the uncertainty of the Weather Regime definitions. 

Second, because of the instability of WR definitions when trained on CMIP6 model data, it should be investigated whether assigning regimes trained on ERA5 reanalysis data to CMIP6 historical data, produces a reliable signal. 
If that is the case, these regimes trained on ERA5 reanalysis data could be assigned to SSP data streams of the CMIP6 models.
That could give insight into the possible future behaviours of regimes under climate change.
This has already been done by \textcite{Fabiano2021-mm}, however, there only the behaviour of the WR4 definition is investigated. 
Furthermore, \textcite{Mller2025} applies the technique of using ERA5 reanalysis data as a reference for the regimes and applies it to CMIP6 model data as well, however, there it is applied on a summertime definition of WR's.


\section{Conclusion}
\label{sec:conclusion}
The aim of this research was to assess the stability of a Weather Regime framework for analysis and its suitability for impact research within the energy sector. 

We carried out an extensive validation of various Weather Regime definitions, however this was hindered by the limited availability of source material. 
Many works do not carry out this vital validation step within their WR framework.
We found that in general the incomplete definitions, the lack of published open data, and the absence of code, means that a strenuous effort is necessary to be able to reliably reproduce WR definitions. 

Our sensitivity assessment showed that the persistence and occurrence of regimes produced by all of the WR definitions we investigated (WR4, WR6 \& WR7) are sensitive to small changes in the methodology. 
Both definition dependent parameters (i.e. the number of EOFs and the method of assignment) and data dependent parameters (i.e. region, time- and spatial resolution, considered months and time period) affect the occurrence and persistence of the regimes.
Though, for each WR definition, Weather Regime and parameter combination, the occurrence and persistence vary with changes in these parameters to a different degree. 

A key sensitivity we identified, is that the stability of the regimes defined with WR4 and WR7 are highly dependent on the period the clustering algorithm is trained on in ERA5. 
As we showed, not all the expected regimes can be identified when the clustering is performed on data from the period 1991--2020.
Therefore, we recommend using periods of at least 50 years, when training WR definitions for climate system dependent implementations, such as the energy sector.

We found that the WR framework is not usable for climate change research as there is no reliable performance within the CMIP6 sub-ensemble analysed.
When applying the WR definitions to the piControl and historical data streams of a sub-ensemble of CMIP6 models, none of the definitions could reliably produce the expected circulation patterns. 
For this we used the normalised distance between the regimes found within CMIP6 and those from ERA5.
The WR4 definition produces the expected regimes for 10/12 of our experiments, the WR6 definition for 0/9 experiments, and the WR7 definition for 9/9 experiments.  

Though, the WR4 and WR7 definitions perform well in terms of number of experiments that pass the threshold, these results must be treated with caution.
The reason for this is the instability of the WR4 and WR7 definitions found within our sensitivity assessment, especially with regards to the time periods in the ERA5 reanalysis data. 
The occurrence and persistence rates found within the CMIP6 sub-ensemble are not reliable.
Our findings imply that CMIP6 models are not suitable for the identification of Weather Regimes and an assessment of climate change effects on them is thus currently unfeasible. 
The use of the WR framework presented with CMIP6 model data is thus unreliable as an indicator of the frequency of future energy droughts or a more general assessment of climate change impact for the energy sector.

Understanding the uncertainties in Weather Regime definitions, the associated impacts of specific WRs, and specific effects of climate change on these Weather Regimes, will help negate possible risks in the energy transition.
Planning for renewable energy infrastructure requires an adequate impact model.
To integrate WR analysis into impact analyses for the energy sector, one definition, is not reliable enough for policy making, because of the large variability present in the definitions.
Using several standardised methods, with a robust understanding of the uncertainties within each method can improve the reliability of the outcome of an impact model. 
For this, more transparency regarding methodology and establishing standard WR definitions is required. 
However for now, these WR definitions are not suitable to be implemented in impact models for the energy transition.

 
\section*{Acknowledgments}
The content of this paper and the views expressed in it are solely the author’s responsibility, and do not necessarily reflect the views of TenneT TSO B.V..

We acknowledge the World Climate Research Programme, which, through its Working Group on Coupled Modelling, coordinated and promoted CMIP6. 
We thank the climate modelling groups for producing and making available their model output, the Earth System Grid Federation (ESGF) for archiving the data and providing access, and the multiple funding agencies who support CMIP6 and ESGF.

This document has been produced in the context of the Copernicus Climate Change Service (C3S).\\
\\
The activities leading to these results have been contracted by the European Centre for Medium-Range Weather Forecasts, operator of C3S on behalf of the European Union (Delegation Agreement signed on 11/11/2014 and Contribution Agreement signed on 22/07/2021). 
All information in this document is provided "as is" and no guarantee or warranty is given that the information is fit for any particular purpose.\\
\\
The users thereof use the information at their sole risk and liability.
For the avoidance of all doubt , the European Commission and the European Centre for Medium - Range Weather Forecasts have no liability in respect of this document, which is merely representing the author's view.

\renewcommand\refname{References}
\newrefcontext[sorting=nty]
\printbibliography

\clearpage


\appendix

\beginsupplement

\begin{appendices}


\section{Open Research}
\label{ap:open}

The ERA5 data used in this study is available at the Climate Data Store via \url{https://www.doi.org/10.24381/cds.adbb2d47} under the CC-BY 4.0 License~\citep{ERA5DataStore,ERA5Data}. 
See SI Section~\ref{method:ERA5} for further details on the ERA5 dataset and the specific parameters used here. 

Table \ref{tab:cmip6models} shows which CMIP6 models are investigated here, with their references. 
See SI Section~\ref{method:CMIP6} for further details on the CMIP6 dataset.

\begin{table}[h]
\centering
\caption{CMIP6 models that are considered to assess the suitability of WR methodology on climate models and their relevant references. }
\label{tab:cmip6models}
\begin{tabular}{l l}
CMIP6 Models                           \\ \hline

\multicolumn{1}{|l|}{CMCC-CM2-SR5}  & \multicolumn{1}{|l|}{\textcite{Cherchi2019}}   \\ \hline
\multicolumn{1}{|l|}{CESM2}    & \multicolumn{1}{|l|}{\textcite{Danabasoglu2020}}        \\ \hline
\multicolumn{1}{|l|}{NorESM2-MM}    & \multicolumn{1}{|l|}{\textcite{Seland2020}}   \\ \hline
\multicolumn{1}{|l|}{NorESM2-LM}    & \multicolumn{1}{|l|}{ \textcite{Seland2020}}      \\ \hline
\multicolumn{1}{|l|}{UKESM1-0-LL}   & \multicolumn{1}{|l|}{\textcite{Sellar2019}}      \\ \hline
\multicolumn{1}{|l|}{MPI-ESM1-2-LR}  & \multicolumn{1}{|l|}{\textcite{Mauritsen2019}}   \\ \hline
\multicolumn{1}{|l|}{EC-Earth3-Veg-LR}  & \multicolumn{1}{|l|}{\textcite{Dscher2022}}  \\ \hline
\end{tabular}
\end{table}

Matteo de Felice's full Weather Regime cookbook, that was implemented in our code,  can be found on \url{https://github.com/matteodefelice/a-recipe-for-weather-regimes}.
The data that was used there can be found at ZENODA via \url{10.5281/zenodo.8384347} with the CC BY-SA 4.0 license~\citep{matteocode2023}.

Falkena's full Weather Regime data used for the categorisation of weather in their study are available upon request from dr. S.K.J. Falkena or the validation Weather Regime definition used here at ZENODO via \url{https://doi.org/10.5281/zenodo.7782226} with the CC BY-SA 4.0 license~\citep{WRDATA}. 
The method describing their creation is presented in~\cite{falkena2020regimes} and the original implemented code can be found on \url{https://github.com/SwindaKJ/Regimes_Public}.

\textbf{\textcolor{red}{AFGESPROKEN DAT DIT OPGELEVERD WORDT NA INLEVEREN THESIS}}.\\
Our code and implementation of the WR framework presented here and additional figures are available at Github via \url{https://github.com/...} with the MIT license. 
The WR assignments as presented in this work is available at Zenodo via \url{https://www.doi.org/10.5281/zenodo} with the CC-BY 4.0 license


\section{ERA5 dataset}
\label{method:ERA5}
For the validation and sensitivity tests the z500 values of the `ERA5 hourly data on pressure levels from 1940 to present' dataset was used. 
The total period of the dataset we used was 1970-01-01 00:00 UTC to 2024-12-31 18:00:00 UTC and the total region was the Euro-Atlantic region of 
90$^\circ$W-40$^\circ$E, 20$^\circ$N-90.0$^\circ$N with a resolution of 0.25$^\circ$.
Depending on the WR definition, a different subset considering the period, months of the year, region, times sampled per day and grid resolution was used. 
These differences are shown in Tables \ref{tab:WR4settings}, \ref{tab:WR6settings} and \ref{tab:WR7settings} for the WR4, WR6 and WR7 definitions respectively. 


\section{The CMIP6 ensemble}
\label{method:CMIP6}
To test suitability of WR definitions for impact assessment, the WR methods (Section~\ref{method:WRdef}) were applied to seven models from the Coupled Model Intercomparison Project phase 6. 
These seven models and their references are shown in Table \ref{tab:cmip6models}.
All these datasets have a longitude-latitude grid, that we regridded to a resolution of $1 ^\circ$, in line with the grid of the baseline regime described in SI Section~\ref{ap:validCMIP6}.
They are subset to the Euro-Atlantic sector of 90$^\circ$W-30$^\circ$E, 20$^\circ$-80$^\circ$N, so for all WR definitions a uniform region is used.
The WR definitions were applied to the piControl and the historical runs of the models.


\section{Settings validation Weather Regime definitions}
\label{ap:settings}

Here we show what model and data parameter settings were used for the validation experiments. 
First the settings of the literature models are compared to the settings of the WR4, WR6 and WR7 models used here. 
Then the settings of the models applied to the CMIP6 models are explained.

\subsection{Settings validation ERA5 reanalysis regimes}
\label{ap:settingsvalera5}
To validate the findings of our models, we need to compare the results to the original methods. 
This can most accurately be done if the methods are as similar as possible. 
The settings of the model and data parameters of our method and the corresponding literature methods are shown in Tables \ref{tab:WR4settings}, \ref{tab:WR6settings} and \ref{tab:WR7settings} for WR4, WR6 and WR7 respectively. 

There are some key differences in our methods and the methods in the literature we based our methods on. 
For WR4 there are two key differences, that can be seen in Table~\ref{tab:WR4settings}.
The first is that we used a grid resolution of $1^\circ$, because of computational considerations.
The second difference is that \textcite{vanderwiel2019regimes} does not describe how the gph anomalies are calculated.
We calculate the anomalies by subtracting the climatology of each day.

\begin{table}[b]
\centering
\caption{Settings Weather Regime identification and assignment for 4 Weather Regimes (WR4) in \textcite{vanderwiel2019regimes} and settings used for model validation in this project.}
\label{tab:WR4settings}
\begin{tabular}{|l|l|l|}
\hline
                       & WR4 \citep{vanderwiel2019regimes} & WR4 validation      \\ \hline
Period                 & 1979--2018                      & 1979--2018           \\ \hline
Months                 & December -- February (DJF)                            & DJF                 \\ \hline
Region                 & 90$^\circ$W-30$^\circ$E, 20$^\circ$-80$^\circ$N            & 90$^\circ$W-30$^\circ$E, 20$^\circ$-80$^\circ$N \\ \hline
Seasonal Cycle Removed & No                             & No                  \\ \hline
Times resolution & 00:00UTC                       & 00:00UTC            \\ \hline
Grid Resolution        & 0.25$^\circ$                         & 1$^\circ$                  \\ \hline
\# EOFs                & 14                             & 14                  \\ \hline
Low-pass filter        & No                             & No                 \\ \hline
Assignment             & Hard Assignment                & Hard Assignment     \\ \hline

Data                 & ERA5            & ERA5
                \\ \hline
\end{tabular}
\end{table}

The largest difference between our method and the method used in ~\citep{falkena2020regimes} is that we use the ERA5 reanalysis dataset and they used the ERA5-interim dataset.
This is shown in Table~\ref{tab:WR6settings}.
Another difference is that to calculate the anomalies \textcite{falkena2020regimes} subtracts a background state, while here the climatology of the each specific day is subtracted.

\begin{table}[t]
\centering
\caption{Settings Weather Regime identification and  assignment for 6 Weather Regimes (WR6) in \textcite{falkena2020regimes} and settings used for model validation in this project.}
\label{tab:WR6settings}
\begin{tabular}{|l|l|l|}
\hline
                       & WR6 \citep{falkena2020regimes} & WR6 validation      \\ \hline
Period                 & 1979--2018                   & 1979--2018           \\ \hline
Months                 &  December -- March (DJFM)                        & DJFM                \\ \hline
Region                 & 90$^\circ$W-30$^\circ$E, 20$^\circ$-80$^\circ$N         & 90$^\circ$W-30$^\circ$E, 20$^\circ$-80$^\circ$N \\ \hline
Seasonal Cycle Removed & No                          & No                  \\ \hline
Times resolution & 00:00UTC                    & 00:00UTC            \\ \hline
Grid Resolution        & 2.5$^\circ$                        & 2.5$^\circ$                \\ \hline
\# EOFs                & n.a.                        & n.a.                \\ \hline
Low-pass filter        & No                        & No               \\ \hline
Assignment             & Hard Assignment             & Hard Assignment     \\ \hline
Data                 & ERA5-Interim                & ERA5
                \\ \hline

\end{tabular}
\end{table}

Settings of our WR7 method differ from the original in four aspects.
First, in \textcite{Grams2017weather} the anomalies are calculated with respect to a 90-day running mean as reference climatology, however, here the average of each day, over the entire time period, is calculated and subtracted from that day.
Secondly, the method of seasonal cycle removal is not the same. 
Where \textcite{Grams2017weather} uses a 30-day rolling window to subtract the seasonal cycle, we subtract the monthly mean of each data point. 
Third, it was not possible to produce a low-pass filter with a reasonable run-time, so the usage of a low-pass filter was discarded. 
Lastly, the method of assignment of the WR-index was not replicated, because we decided to move forward with a Bayesian assignment method.

\begin{table}[t]
\centering
\caption{Settings Weather Regime (WR) identification and assignment for WR7 in \textcite{Grams2017weather} and settings used for model validation in this project.}
\label{tab:WR7settings}
\begin{tabular}{|l|l|l|}
\hline
                       & WR7 \citep{Grams2017weather}   & WR7 validation              \\ \hline
Period                 & 1979--2015                   & 1979--2015                   \\ \hline
Months                 & Full year                   & Full year                   \\ \hline
Region                 & 80$^\circ$W-40$^\circ$E, 30$^\circ$-90$^\circ$N         & 80$^\circ$W-40$^\circ$E, 30$^\circ$-90$^\circ$N         \\ \hline
Seasonal Cycle Removed & Yes, 30 day rolling window  & Yes, monthly mean removed   \\ \hline
Times resolution & \small 00:00,06:00,12:00,18:00 UTC & \small 00:00,06:00,12:00,18:00 UTC \\ \hline
Grid Resolution        & 1$^\circ$                          & 1$^\circ$                         \\ \hline
\# EOFs                & 7                           & 7                           \\ \hline
Low-pass filter        & 10 days                     & No                          \\ \hline
Assignment             & WR Index                    & Hard Assignment             \\ \hline
Data                 & ERA5-Interim               & ERA5
                \\ \hline
\end{tabular}
\end{table}

\subsection{Settings validation CMIP6 regimes}
\label{ap:validCMIP6}
When applying the WR definitions to CMIP6 models, the settings are as close to the original literature settings as possible, as described previously in SI Section~\ref{ap:settingsvalera5}.
However, there are four notable differences. 
The first difference applies to WR4, WR6 and WR7:
The considered period of the data is for all models the complete available dataset. 
The second difference also applies to all three definitions and involves the grid resolution. 
Each model has a unique grid with a specific resolution, but the data is regridded to a $1^\circ$ grid, based on the grid of the ERA5 reanalysis data. 
The third difference is that for WR7, the considered region is changed to 90$^\circ$W-30$^\circ$E, 20$^\circ$-80$^\circ$N. 
The last difference is that for the WR7 definition the time resolution is reduced to one data point per day. 

When comparing the regimes found using CMIP6 model data to regimes found using ERA5 reanalysis data, as described in SI Section \ref{ap:regimesautomatic}, the ERA5 reanalysis regimes are found with `baseline' settings that are the same for each definition.
These settings can be found in Table \ref{tab:CMIP6ERA5baseline}.
The definition specific settings, number of EOFs and which months are included, do differ between definitions.

\begin{table}[h]
\centering
\caption{Baseline settings used for the baseline regimes, found with ERA5 reanalysis data, with the goal of comparing regimes found with the ERA5 reanalysis data and regimes found with CMIP6 experiment data.
The CMIP6 experiment data uses the same settings, with the exception of the period, there the entire available period is used.}
\label{tab:CMIP6ERA5baseline}
\begin{tabular}{|l|
>{}l |}
\hline
Period          & 1971--2020                                                              \\ \hline
Grid resolution & { $1^\circ$}                                       \\ \hline
Region          & { 90$^\circ$W-30$^\circ$E, 20$^\circ$-80$^\circ$N} \\ \hline
Time resolution & 00:00UTC                                                               \\ \hline
\end{tabular}
\end{table}


\section{Weather Regime Circulation patterns}
\label{ap:figureswr}
Here we will discuss how the Weather Regimes (WR) differ from previous research and from eachother.
First we will establish the naming conventions we used in this paper to describe the Weather Regimes (WR).
Second we show how the circulation patterns from WR4 and WR6 differ from the WR7 definition.
In the third subsection we discuss how the occurrence and persistence rates of the regimes we found for each definition differs from the corresponding literature values. 
Then we show two examples of the geopotential height (gph) fields produced by applying WR definitions on CMIP6 data.
Finally, we explain how the automatic assignment of regimes to CMIP6 experiment clusters was done. 

\subsection{Naming conventions}
\label{ap:naming}
In this paper, the names of the WR's from different WR definitions are standardized for consistency. 

WR4 and WR6 use the same terminology, except, because no negative phase of AR+ and SB+ exists in WR4, no +/- indication is given in the original work~\citep{vanderwiel2019regimes}.
Here we will refer to AR as AR+, and SB as SB+, so it is equivalent to the terminology of the other regimes.

Table \ref{tab:WRnames} includes a list of the names of the Weather Regimes used here and their equivalent used in \textcite{Grams2017weather}, on which the WR7 definition is based. 

\begin{table}[th!]
\centering
\caption{Names Weather Regimes used in this paper and their equivalent name in Grams(2017)}
\label{tab:WRnames}
\begin{tabular}{l l}
Name Weather Regime (WR)                                  & Equivalent WR in Grams (2017)                     \\ \hline
\multicolumn{1}{|l|}{North Atlantic Oscillation + (NAO+)} & \multicolumn{1}{l|}{Zonal regime (ZO)}      \\ \hline
\multicolumn{1}{|l|}{North Atlantic Oscillation - (NAO-)} & \multicolumn{1}{l|}{Greenland Blocking (GL)}            \\ \hline
\multicolumn{1}{|l|}{Scandinavian Blocking + (SB+)}       & \multicolumn{1}{l|}{Scandinavian Blocking (ScBL)} \\ \hline
\multicolumn{1}{|l|}{Scandinavian Blocking - (SB-)}       & \multicolumn{1}{l|}{Scandinavian Trough (ScTr)}   \\ \hline
\multicolumn{1}{|l|}{Atlantic Ridge + (AR+)}              & \multicolumn{1}{l|}{Atlantic Ridge (AR)}          \\ \hline
\multicolumn{1}{|l|}{Atlantic Ridge - (AR-)}              & \multicolumn{1}{l|}{Atlantic Trough (AT)}         \\ \hline
\multicolumn{1}{|l|}{European Blocking (EB+)}             & \multicolumn{1}{l|}{European Blocking (EuBL)}     \\ \hline
\end{tabular}
\end{table}

\subsection{WR4 and WR6 circulation patterns}
\label{ap:wr4wr6patterns} 
Here we show the regimes found by applying the WR4 and WR6 methods described in \ref{method:WRdef}.
We also point out the visual differences between the WR7 definition's regimes shown in Figure~\ref{fig:WRvisual}.

\begin{figure}[th!]
    \centering
    \includegraphics[width=0.8\linewidth]{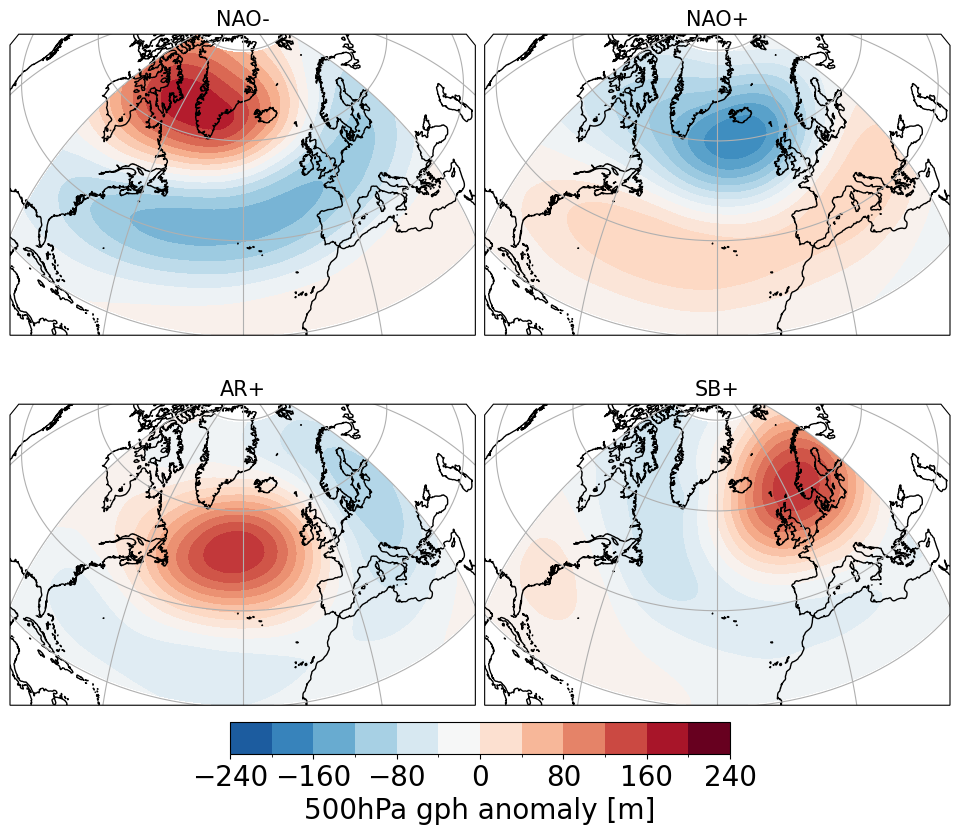}
    \caption{
    As Figure~\ref{fig:WRvisual}, but for the WR4 definition based on the methodology described in Section \ref{method:WRdef} with the settings shown in Table \ref{tab:WR4settings}.}
    \label{fig:WR4base}
\end{figure}

Figure \ref{fig:WR4base} shows the gph fields produced by applying the WR4 definition.
Compared to the WR7 regimes, as shown in Figure \ref{fig:WRvisual}, the WR4 AR+ and SB+ high pressure areas lie more Southward, and the NAO- high pressure area is a smaller anomaly.

\begin{figure}[hb!]
    \centering
    \includegraphics[width=0.8\linewidth]{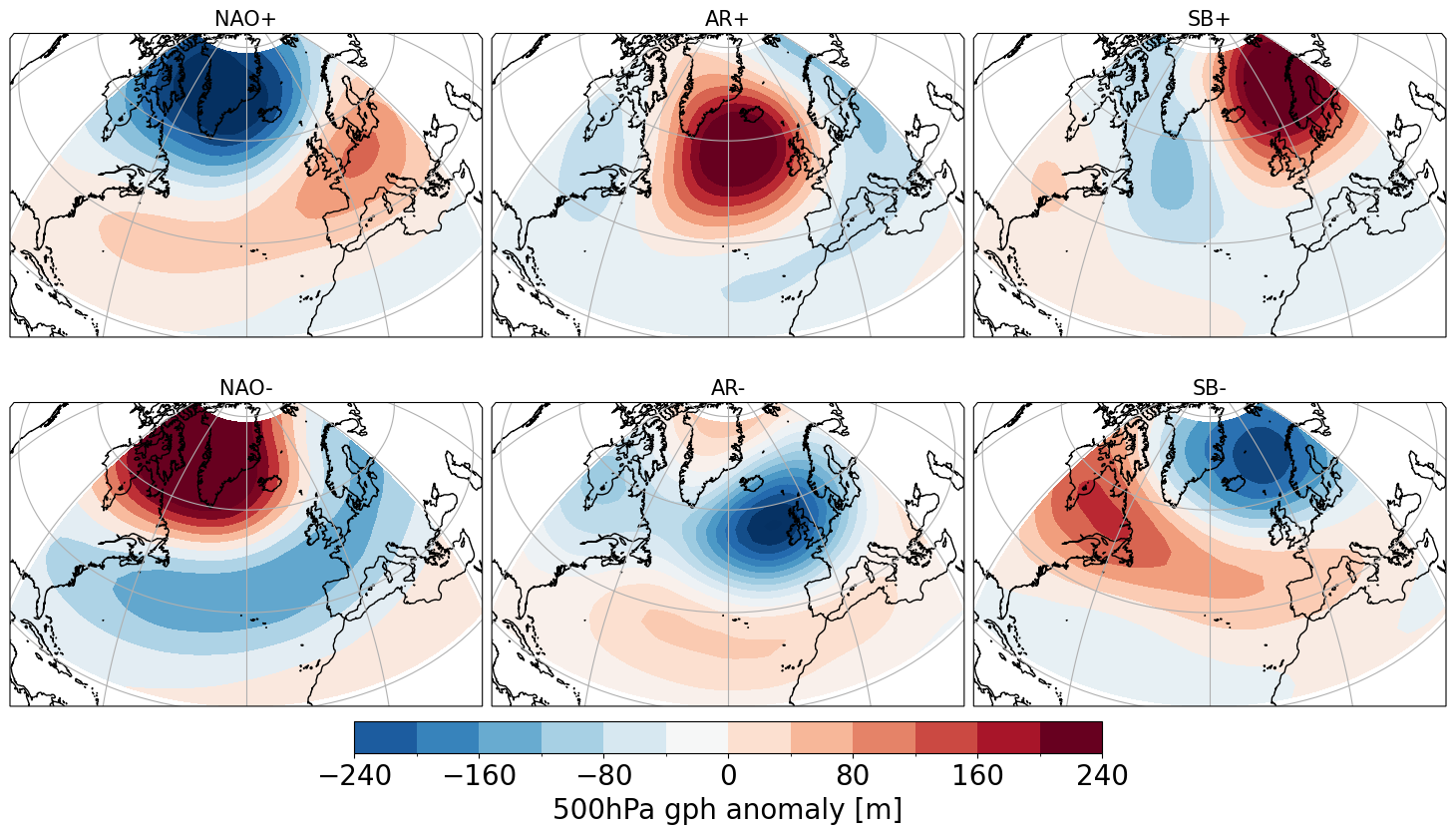}
    \caption{
    As Figure~\ref{fig:WRvisual}, but for the WR6 definition based on the methodology described in Section \ref{method:WRdef} with the settings shown in Table \ref{tab:WR6settings}.}
    \label{fig:WR6base}
\end{figure}

Shown in Figure \ref{fig:WR6base} are the gph fields found by using the WR6 definition.
Generally the WR6 definition has higher anomalies and encompassed larger areas for each regime compared to the WR7 regimes. 
Moreover, the AR- low pressure area is more Eastward for the WR6 definition compared to the WR7 definition. 

\clearpage

\subsection{Occurrence and persistence rates validation}
\label{ap:validoccpers}
In Section \ref{res:validation} of the main text we discuss the occurrence and persistence of the WR definitions.
Here we show the exact persistence and occurrence rates for the regimes produced by applying the WR4 and WR6 definitions with the settings shown in Section \ref{ap:settings}.
We then compare these to the values in the literature, these definitions were based on.

\begin{table}[th!]
\centering
 \caption{Occurrence (Occ) and persistence (Per) rates of the validation experiment for the WR4 definition compared to the values from \textcite{vanderwiel2019regimes}. Please note that no persistance of the WR4 was provided by the reference paper. }\label{tab:validationWR4}
\begin{tabular}{|l|r|r|r|}
\hline
WR   & Occ & Occ literature & Per (days) \\ \hline
NAO+ & 32.1\%     & 33\%                  & 6.0                \\ \hline
NAO- & 19.4\%     & 20\%                  & 6.4                \\ \hline
SB+   & 27.6\%     & 28\%                  & 4.8                \\ \hline
AR+   & 21.0\%     & 20\%                  & 4.1               \\ \hline
\end{tabular}
\end{table}

As can be seen in Table \ref{tab:validationWR4}, we find that, occurrence rates of weather regimes for the WR4 regimes are similar to the values in ~\citep{vanderwiel2019regimes}, but not identical.
There is at most a $1\%$ difference in occurrence rate.
Persistence rates were not available in ~\citep{vanderwiel2019regimes}, so these cannot be compared. 

SI Table \ref{tab:validationWR6} shows that for the WR6 definition we find a persistence that is at most 0.6 days different from the values found in ~\textcite{falkena2020regimes}. 
Furthermore, this Table shows that the difference of occurrence rate of the regimes we calculate and the regimes from ~\citep{falkena2020regimes} can be up to $2.2 \%$. 

It is not possible to compare occurrence and persistence rates for the WR7 definition. 
Occurrence rates cannot be compared, because we could not replicate the `no regime' regime from ~\citep{Grams2017weather}. 
Taking out one regime automatically alters the occurrence rates of the other regimes, therefore it cannot be compared.
Persistence rates cannot be compared because they were not given in ~\citep{Grams2017weather}.
The persistence and occurrence rates we found for WR7 can be seen in Table~\ref{tab:validationWR7}.

\begin{table}[hb]
\centering
 \caption{Occurrence (Occ) and persistence (Per) rates of the validation experiment for the WR6 definition compared to the available literature values from \textcite{falkena2020regimes}. }\label{tab:validationWR6}
\begin{tabular}{|l|r|r|r|r|}
\hline
WR   & Occ & Occ literature & Per (days) & Per literature (days) \\ \hline
NAO+ & 18.1\%     & 16.9\%                & 4.1                & 4.0                           \\ \hline
NAO- & 14.1\%     & 15.5\%                & 5.9                & 6.5                           \\ \hline
SB+  & 19.3\%     & 19.6\%                & 4.2                & 4.0                           \\ \hline
SB-  & 13.9\%     & 16.1\%                & 4.1               & 3.7                           \\ \hline
AR+  & 15.6\%     & 15.6\%                & 3.3             & 3.5                           \\ \hline
AR-  & 18.0\%     & 16.3\%                & 4.1               & 4.7                           \\ \hline
\end{tabular}
\end{table}

\begin{table}[ht]
\centering
 \caption{Occurrence (Occ) and persistence (Per) rates of the validation experiment for the WR7 definition.}
 \label{tab:validationWR7}
\begin{tabular}{|l|l|l|}
\hline
WR   & Occ                           & Per(days) \\ \hline
NAO+ & {\color[HTML]{000000} 14.2\%} & 3.1 \\ \hline
NAO- & {\color[HTML]{000000} 8.6\%}  & 3.6 \\ \hline
SB+  & 16.8\%                        & 3.1 \\ \hline
SB-  & 16.5\%                                                & 2.9                         \\ \hline
AR+  & 10.8\%                        & 3.0 \\ \hline
AR-  & 15.6\%                        & 2.7 \\ \hline

EB+  & 17.5\%                                                & 2.7                         \\ \hline
\end{tabular}
\end{table}

\subsection{CMIP6 experiment regimes}
\label{ap:regimesCMIP6}
Here we show two examples of CMIP6 experiments we have executed, to visualise how they are different from the expected regimes.

\begin{figure}[p]
    \centering
    \includegraphics[width=0.8\linewidth]{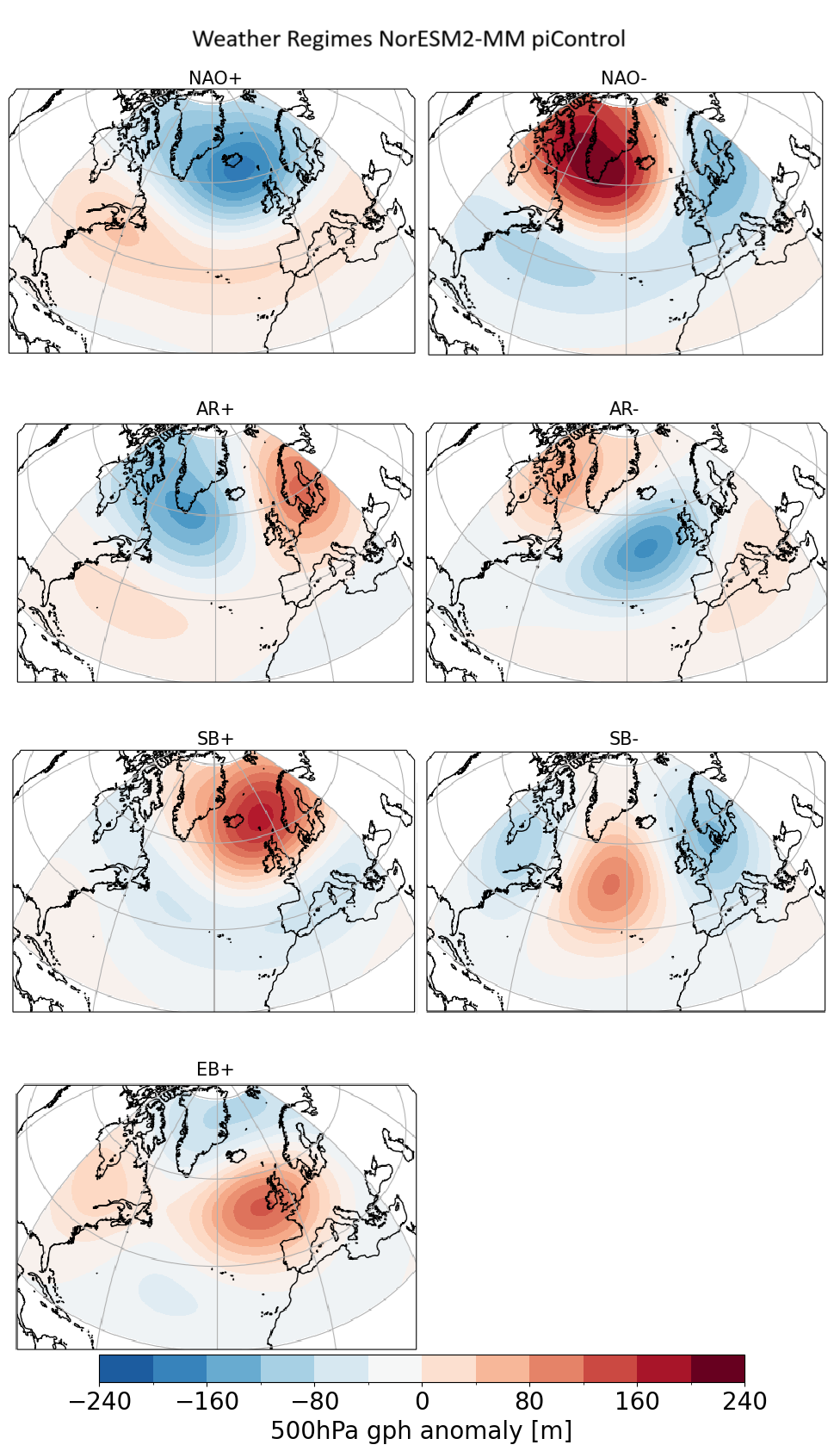}
    \caption{As Figure \ref{fig:WRvisual}, but here the WR7 definition is used to cluster the NorESM2-MM piControl data from the CMIP6 model ensemble. }
    \label{fig:NorESM2piControlWR7}
\end{figure}

First we will discuss the regimes produces by applying the WR7 definition to the NorESM2-MM Picontrol experiment.
This experiment passes the normalised distance threshold as described in \ref{ap:regimesautomatic}, but visually, the regimes are different from the regimes in Figure \ref{fig:WRvisual}.
The regimes produces by the NorESM2-MM Picontrol experiment are shown in Figure \ref{fig:NorESM2piControlWR7}.
There it can be seen that is no clear AR+, but instead a regime resembling SB+ is found. 
Additionally, the EB+ regime is located further towards the Atlantic Ocean than in Figure \ref{fig:WRvisual}.
Similar behaviour, the shifting of regimes or not finding a regime clearly, is present in three other experiments for the WR7 definition. 
These experiments are: NorESM2-MM historical, MPI-ESM1-2-LR historical and UKESM1-0-LL historical.

\begin{figure}[ht]
    \centering
    \includegraphics[width=\linewidth]{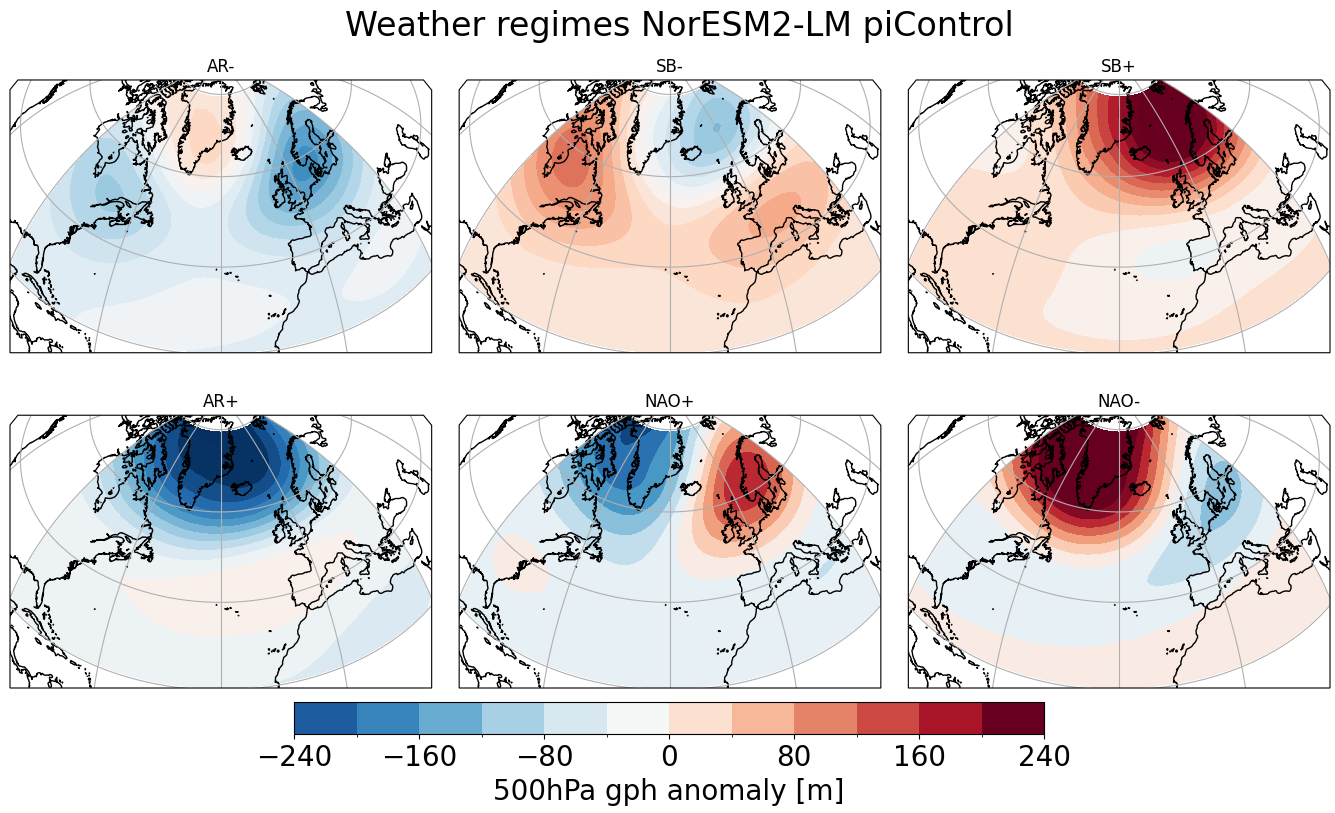}
    \caption{
    Figure as \ref{fig:WRvisual}, but instead of the WR7 definition applied to ERA5 reanalysis data, here the WR6 definition clusters the piControl stream of NorESM2-LM. 
    No AR- regime can be identified.
    }
    \label{fig:WR6CMIP6NorLM}
\end{figure}

Now we discuss the results of the NorESM2-LM piControl experiment, as an example of an experiment that does not pass the normalised distance threshold.
Figure \ref{fig:WR6CMIP6NorLM} shows the regimes produced by training the WR6 definition on the NorESM2-LM piControl data. 
Here some regimes are the regimes we expect to see, like the NAO- and the regime labeled as NAO+ by the matching algorithm resembles SB+. 
However, the other regimes have vastly different structures compared to SI Figure \ref{fig:WR6base}.
For instance, there is no regime with an intense low or high over the central Atlantic Ocean, as is expected of the AR- and AR+ regimes. 

For most CMIP6 experiments that do not pass the normalised distance threshold, some of the expected patterns can be identified, but others are similar to other regimes that have shifted or they are completely different patterns.
For WR6 the regime that is absent most often is the SB- regime.
However, the NAO- and NAO+ are reliably found by the WR6 definition across models and data sets. 
For WR4 AR+ is missing for CMCC-CM2-SR5 piControl, but for NorESM2-LM piControl all regimes have a different pattern to the expected regime. 

\subsection{Automatic matching of regimes}
\label{ap:regimesautomatic}

Identifying regimes by eye, for each experiment, is tedious and prone to human error. 
So for the experiments using the CMIP6 data, we applied automatic matching based on the distance between the new regime and a so-called baseline regime (For the ERA5-reanalysis sensitivity tests, the matching was done visually by hand for each experiment). 
This process consists of five steps:
First a clustering was done on ERA5 data with the settings shown in Table~\ref{tab:CMIP6ERA5baseline}. 
These clusters are then our baseline regimes.
Second, the gph anomalies of the CMIP6 experiment were regridded towards the grid of the ERA5 clusters, using an interpolation algorithm. 
Third, the regridded gph anomaly data from the CMIP6 experiment (with the settings of Table \ref{tab:CMIP6ERA5baseline}, with the exception of the time period.) was processed according to the corresponding WR definition method described in Section~\ref{method:WRdef}. 
Fourth, the distance between the clusters found in the CMIP6 experiment data and the clusters from the ERA5 baseline data was calculated.
This is done for each combination of clusters.
Lastly, the cluster centres are matched such that the sum of the distances between the ERA5 and CMIP6 results is minimized. This is particularly relevant when there is not a one-to-one match between the cluster centres, with for example two CMIP6 clusters sharing an ERA5 cluster they are closest to. The minimization optimizes the match, but it will result in CMIP6 clusters not always being assigned to their closest ERA5 cluster.

The smallest distances, one distance for each cluster, are then added to each other and normalised. 
We then use this normalised distance to assess how successful the clustering of the CMIP6 experiment is. 
If the normalised distance is smaller than 0.4 we consider the CMIP6 experiment regimes to be similar to the ERA5 baseline regimes.


\newpage
\section{Sensitivity}
\label{ap:figuressens}
As we discussed in Section~\ref{res:sensitivity} of the main text, we evaluated the sensitivity of WR definitions to altering parameters.
Here we show the results of the parameter sensitivity tests of the WR4, WR6 and WR7 definitions on ERA5 reanalysis data in more detail and with additional background. 
First we discuss how altering parameters inherent to the WR definition, affects the persistence and occurrence of the regimes within those definitions. 
Then we discuss how changing data parameters affects the regimes. 
There we will show changes in occurrence and persistence and highlight how the WR7 regimes look when the period 1991--2020 is used for clustering. 

\subsection{WR definition parameter sensitivity}
\label{ap:senswrdefparam}
There are two types of model parameters and how they affect the occurrence and persistence, namely the number of EOFs and the method of assignment. 

First the number of EOFs.
For WR4 and WR7 the number of EOFs used in the first step of calculating the regimes was varied between 7, 14 and 30 EOFs.
This was done, because choosing the wrong number of EOFs can either over or underfit the data.
So, understanding how the number of EOFs influences the persistence or occurrence of certain regimes, can inform which number would be best practice. 
Second, the effect of changing the method of assignment, hard or Bayesian assignment, is investigated.
The assignment of each data point to the closest cluster is a straightforward, computationally light, technique.
However, it is not a realistic representation of the atmospheric circulation, because WRs represent persistent patterns.
These do not change daily.
Therefore, we also looked into the effect of using a probabilistic method of assignment, as this results in more gradual transitions.

\begin{figure}[ht!]
    \centering
    \includegraphics[width=\linewidth]{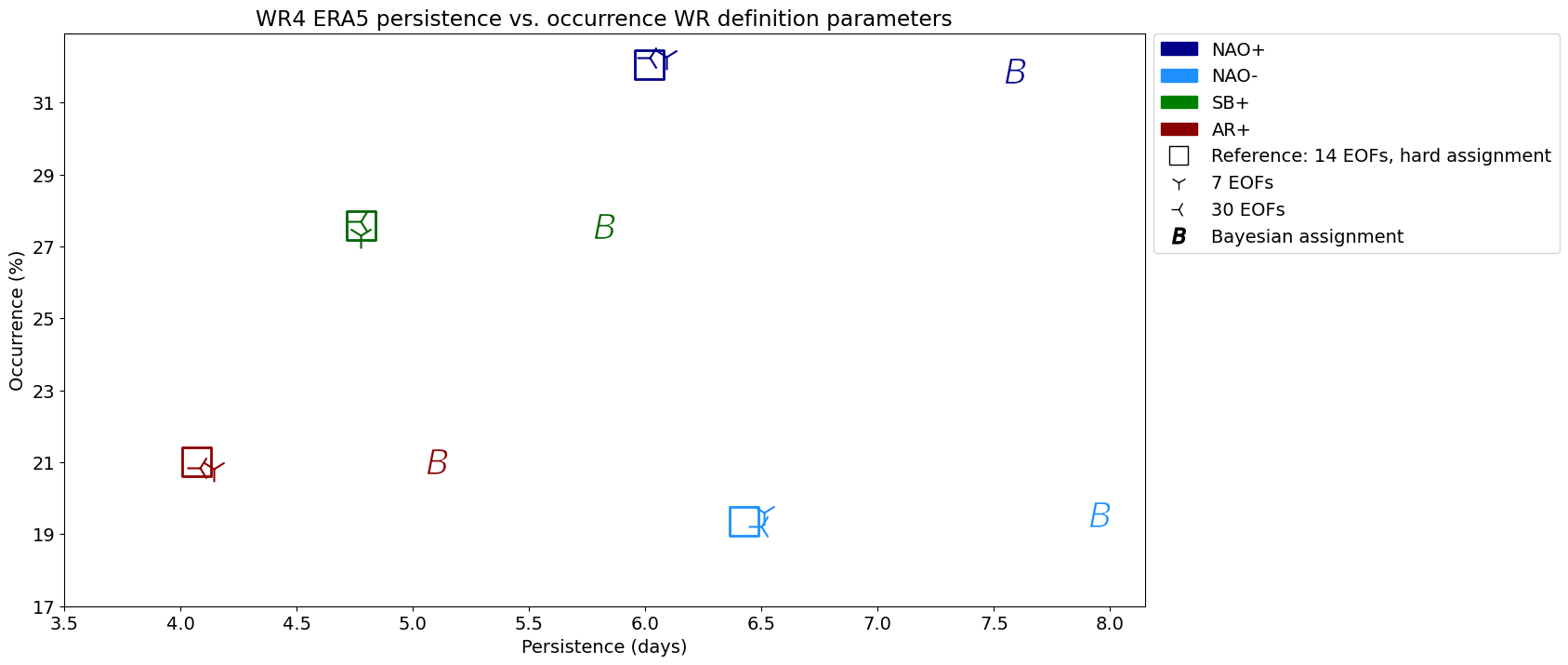}
    \caption{
    As Figure~\ref{fig:WR4periods}, but for the varied WR definition parameters: 
    The number of EOFs and method of assignment
    7 or 30 EOFs means that for that run, 14 EOFs was changed to 7 or 30 EOFs. 
    Bayesian assignment means, that instead of hard assignment the method of assignment is based on 
~\citep{FalkenaBayes2023}.
   Note that the x-axis is not identical to the x-axis in Figure~\ref{fig:WR4periods}.}
    \label{fig:WR4sensmodel}
\end{figure}


Figure \ref{fig:WR4sensmodel} shows that changing the number of EOFs does not have a large effect for the WR4 definition. 
However, for the WR7 definition, Figure \ref{fig:WR7sensmodel} indicates that a larger number of EOFs increases the persistence for some regimes.
The difference between 14 and 30 EOFs is small, so the largest difference is between 7 and 14 EOFs. 
This implies that the leading 7 EOFs do not capture all the relevant variance.

\begin{figure}[hb!]
    \centering
    \includegraphics[width=0.8\linewidth]{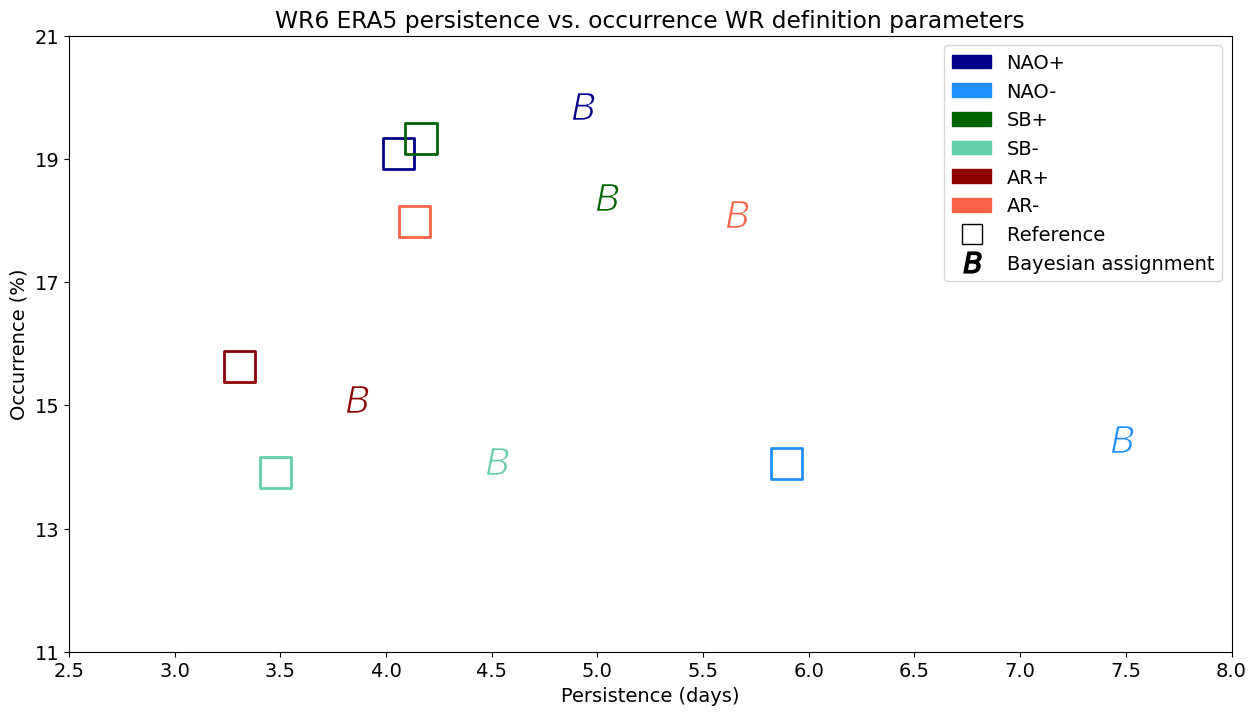}
    \caption{As Figure\ref{fig:WR4periods}, but instead of the WR4 definition, here the WR6 definition~\citep{falkena2020regimes} is applied and the WR definition parameters are varied.
    For the WR6 definition that means the hard assignment is changed to Bayesian assignment as explained in Section~\ref{method:WRdef}.
    The abbreviations of the regimes are explained in Section \ref{method:WRdef}.}
    \label{fig:WR6sensmodel}
\end{figure}

Changing the method of assignment from hard assignment to Bayesian assignment increases the persistence for all regimes for each WR definition. 
This change affects the occurrence rate per regime differently for each WR definition. 
For WR4 only a small change in occurrence can be identified, as seen in Figure \ref{fig:WR4sensmodel}. 
Figure \ref{fig:WR6sensmodel} shows that for WR6, an increased occurrence for NAO+ and NAO- and a decreased occurrence for SB+ and AR+ can be identified. 
WR7 also shows an increased occurrence for NAO+ and NAO-, but a decreased occurrence for SB+, SB- and AR-, which is shown in Figure \ref{fig:WR7sensmodel}.
Note that, when calculating the persistence of the regimes while using this method, the WR with the highest probability is assigned to each data point.
This is done, to allow us to compare the persistence to the hard method, as it is not possible to calculate a persistence based on a probabilistic distribution of each regime per day in a straight-forward manner.

\begin{figure}[ht!]
    \centering
    \includegraphics[width=\linewidth]{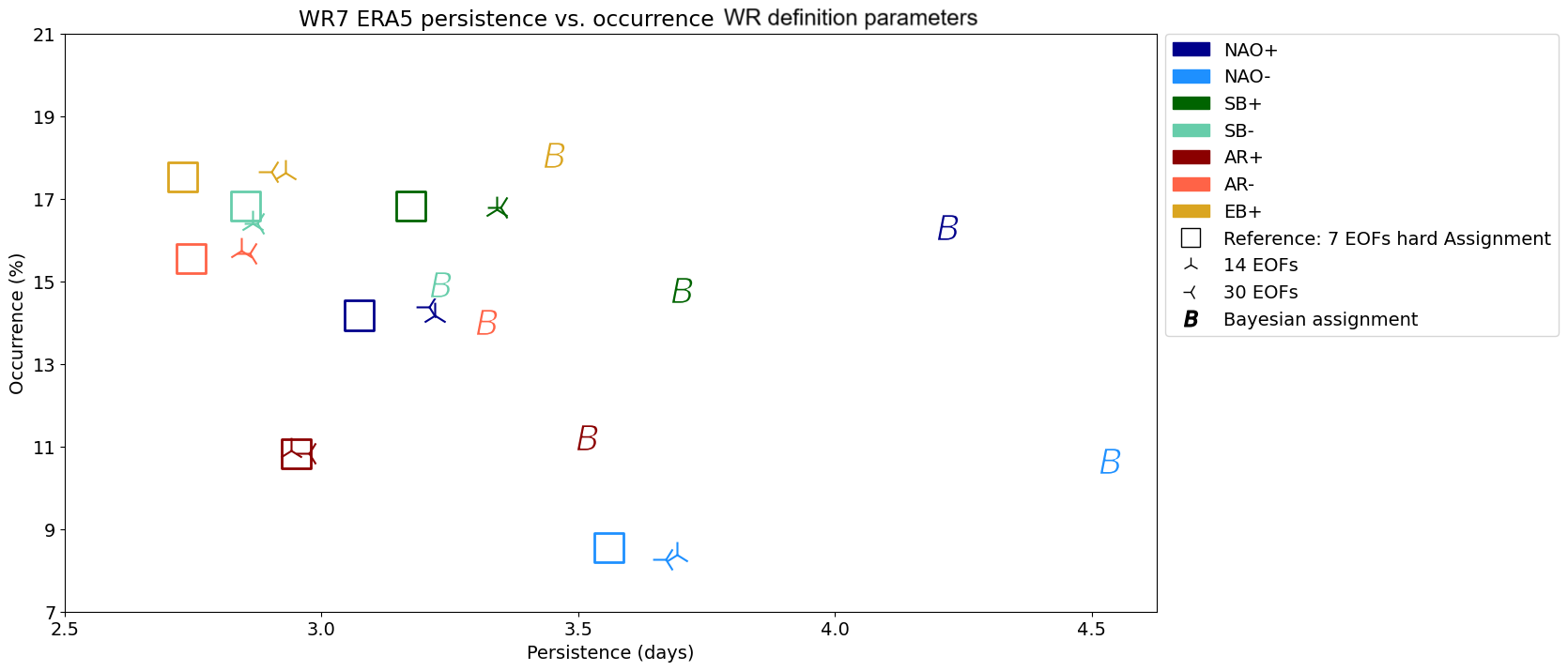}
    \caption{
    As Figure \ref{fig:WR4periods}, but here the WR7 definition~\citep{Grams2017weather} is applied and the WR definition parameters are varied. 
    Those parameters are the number of EOFs and the method of assignment.
    14 or 30 EOFs means that for that run, 7 EOFs was changed to 14 or 30 EOFs. 
    Bayesian assignment means, that instead of hard assignment the method of assignment is based on 
    \textcite{FalkenaBayes2023}.
    The abbreviations of the regimes are explained in Section \ref{method:WRdef}.}
    \label{fig:WR7sensmodel}
\end{figure}

\subsection{Data parameter sensitivity}
\label{ap:sensdataparam}
Data parameters that we have varied can be divided into two categories:
Time dependent parameters and space dependent parameters.

The time dependent parameters we consider are the time period, time resolution and which months are taken into account. 
We look at the time period to understand the stability of the regime definitions across decades.
This is done with two different approaches:
The first approach is to both cluster and assign the data on the same time period. 
Another approach we use is to cluster the data on the data from 1971--2020, the full available period, and then assign the clusters to data in the climatological periods 1971--2000, 1981--2010 and 1991--2020. 
We also research the effects of altering the time resolution, 6-hourly or daily data, to assess whether or not more data translates into a more accurate representation of reality. 
Furthermore, which winter months are used varies between the WR4 and WR6 definitions.
So we investigate if there is a notable difference to persistence or occurrence of certain regimes if in addition to December, January and February being considered, March is included.

Space parameters that are investigated are the grid resolution and the region. 
Because grid resolution greatly affects computational costs, the effect of changing the grid resolution from $1^\circ$ to $2.5^\circ$ can vastly reduce these costs. 
Therefore, understanding if reducing the grid resolution has an influence on the quality of the results is important. 
Euro-Atlantic WR's represent the dominant circulation patterns over the Euro-Atlantic region.
So, we need to understand how defining this region differently affects the geographic location of low- and high-pressure areas within the regimes and how this affects the persistence and occurrence of these regimes within the model.
Therefore we vary the region of the data the regimes are clustered on between the region used in ~\citep{vanderwiel2019regimes}
(90$^\circ$W-30$^\circ$E, 20$^\circ$-80$^\circ$N), the region used in ~\citep{Grams2017weather} (80$^\circ$W-40$^\circ$E, 30$^\circ$-90$^\circ$N), and the following three regions:
90$^\circ$W-40$^\circ$E, 20$^\circ$-90$^\circ$N  
80$^\circ$W-30$^\circ$E, 20$^\circ$-80$^\circ$N
90$^\circ$W-40$^\circ$E, 20$^\circ$-80$^\circ$N.

\begin{figure}[bh!]
    \centering
    \includegraphics[width=\linewidth]{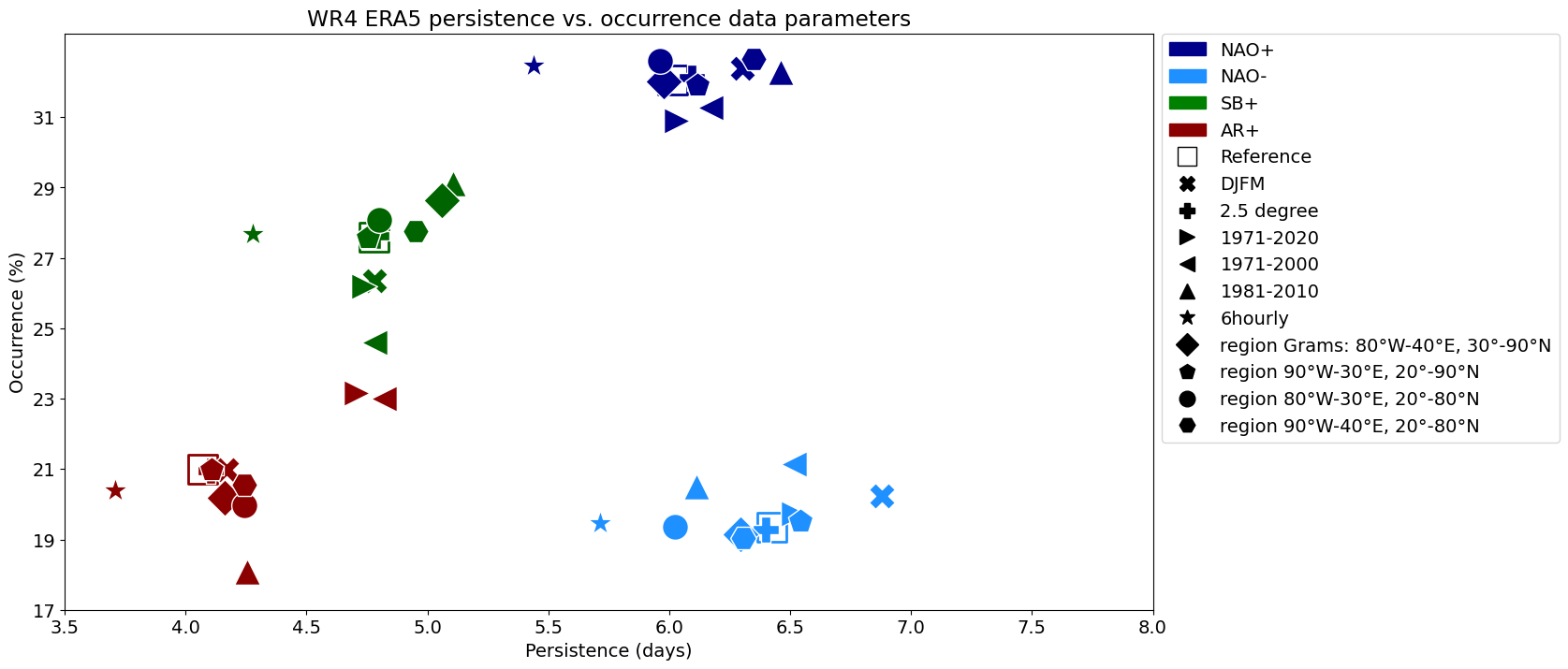}
    \caption{
    As Figure~\ref{fig:WR4periods}, but with additional varied data parameters and excluding the persistence and occurrence calculated by training the clustering on 1971 and assigning it on another decade.
    `DJFM' means that the months taken into account are December, January, February and March. 
    `2.5 degree' means that the grid resolution is 2.5 $^\circ$. 
    The triangles show which period the clustering algorithm is trained on. 
    `6 hourly' means that instead of daily data, 6-hourly data is used. 
    Whenever the legend says `region', the bounds of the region are altered.
    Note that the x-axis here is not identical to that in Figure~\ref{fig:WR4periods} and certain identical shapes indicate a different parameter here.
    }
    \label{fig:WR4sensdata}
\end{figure}

\begin{figure}[ht]
    \centering
    \includegraphics[width=0.9\linewidth]{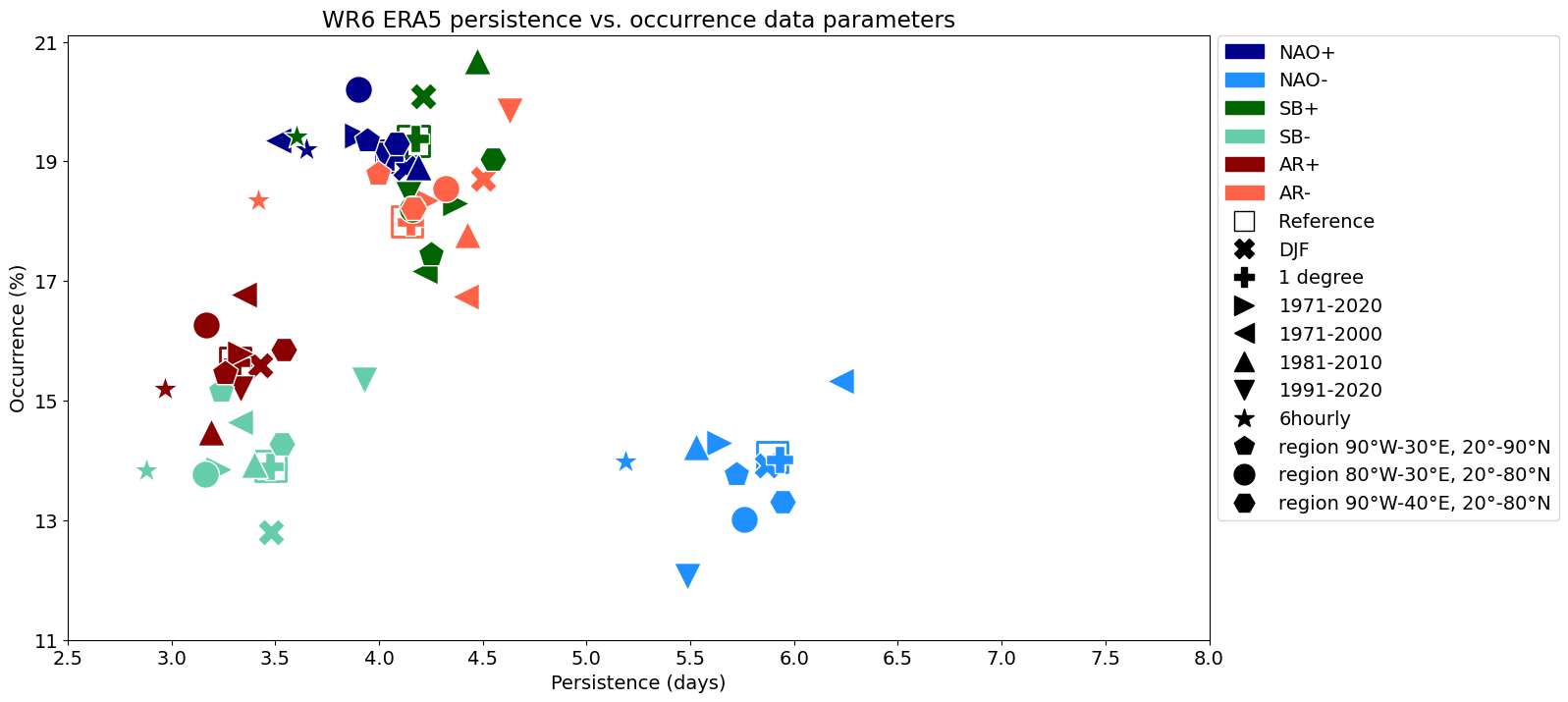}
    \caption{
    As Figure~\ref{fig:WR4periods}, but here the WR6 definition~\citep{falkena2020regimes} is applied and other data parameters, in addition to the period, are varied.
    The shapes show how the run deviates from the reference settings, as shown in SI Table~\ref{tab:WR6settings}. 
    `DJF' means that the months taken into account are December, January and February, so March is excluded.
    `1 degree' means that the grid resolution is 1 $^\circ$ instead of 2.5 $^\circ$. 
    The triangles show which period the clustering algorithm is trained on. 
    `6 hourly' means that instead of daily data, 6-hourly data is used. 
    Whenever the legend says `region', the bounds of the region are altered. 
    The abbreviations of the regimes are explained in Section \ref{method:WRdef}.
    Note that certain shapes present in Figure~\ref{fig:WR4periods} indicate a different parameter here.
}
    \label{fig:WR6sensdata}
\end{figure}

\begin{figure}[hb!]
    \centering
    \includegraphics[width=\linewidth]{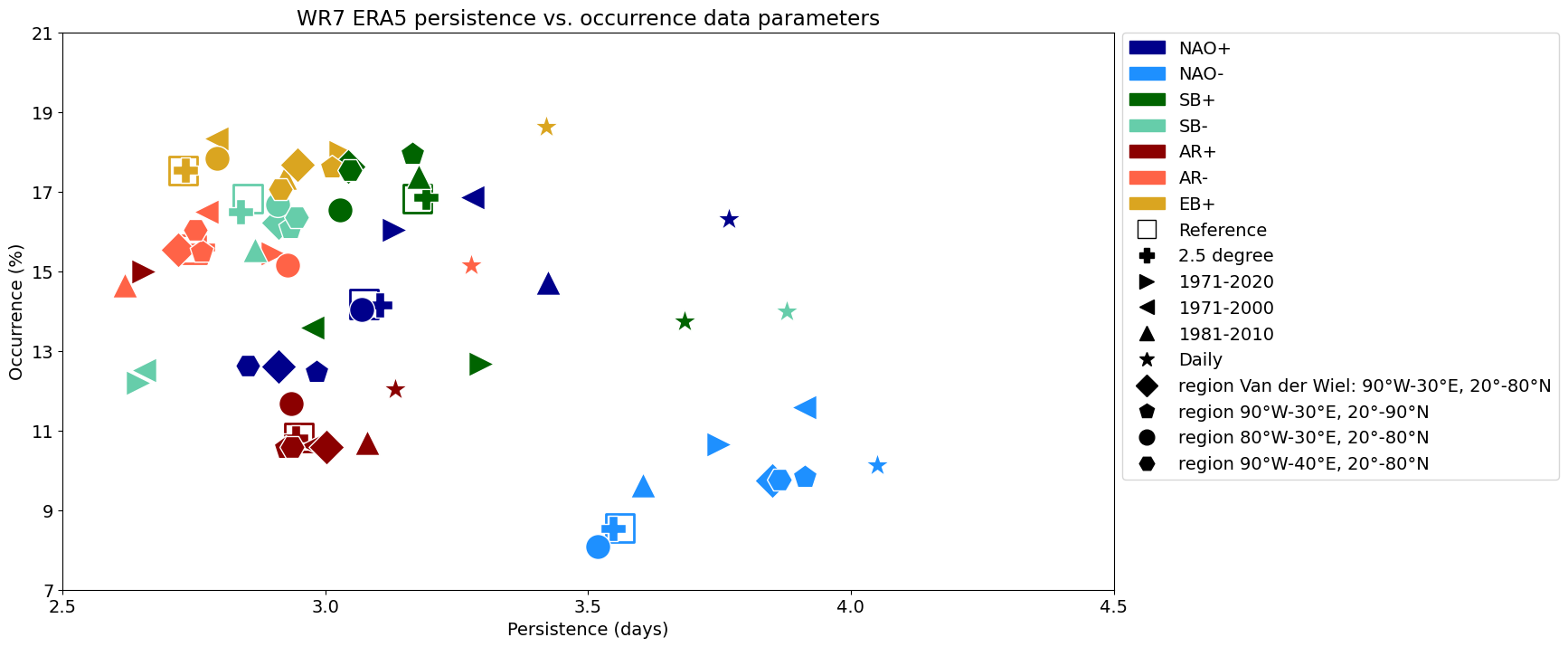}
    \caption{
    As Figure~\ref{fig:WR4periods}, but here the WR7 definition~\citep{Grams2017weather} is applied to varied data parameters. 
    The shapes show how the run deviates from the reference settings, as shown in SI Table~\ref{tab:WR7settings}. 
    `2.5 degree' means that the grid resolution is 2.5 $^\circ$ instead of 1 $^\circ$. 
    The triangles show which period the clustering algorithm is trained on. 
    `Daily' means that instead of 6-hourly data, daily data is used. 
    Whenever the legend says `region', the bounds of the region are altered. 
    The abbreviations of the regimes are explained in Section \ref{method:WRdef}.
    Note that certain shapes present in Figure~\ref{fig:WR4periods} indicate a different parameter here.
}
    \label{fig:WR7sensdata}
\end{figure}

Figures that include the impact of the data parameters, are included in for each WR definition: 
Figures \ref{fig:WR4sensdata}, \ref{fig:WR6sensdata} and \ref{fig:WR7sensdata} for WR4, WR6 and WR7, respectively. 
Those figures show that grid resolution does not have a large effect on either persistence or occurrence.
Furthermore, the temporal resolution only affects the persistence, which is to be expected when using hard assignment. 
More data points means a larger chance to switch to a different regime faster. 
So, for all WR definitions and all regimes the persistence is decreased when switching from daily to 6-hourly data. 
In addition to that, changing the winter months from December--February to December--March, has some small, but noticeable effect for some regimes. 
For the WR4 definition, NAO- has a higher occurrence and longer persistence when March is included and SB+ has a lower occurrence. 
However, for the WR6 definition, the difference for NAO- is negligible when March is excluded.
Here the regimes SB+, SB- and AR- have the most notable shift in occurrence and/or persistence. 

Moreover, the extend of the region does affect the persistence and occurrence of each of the regimes for all WR definitions.

For the WR4 definition Figure \ref{fig:WR4sensdata} shows that shifting the region to 80$^\circ$W-40$^\circ$E, 30$^\circ$-90$^\circ$N, affects the SB+ regime the most, where the persistence and occurrence increase. 
AR+ has a slightly lower occurrence for this region. 
When investigating the effect of altering just the upper bound to 90$^\circ$N, no large change in persistence or occurrence is seen.
Shifting the region Eastward 80$^\circ$W-30$^\circ$E, 20$^\circ$-80$^\circ$N, has the largest effect on NAO-. 
Extending the region Eastward with the region of 90$^\circ$W-40$^\circ$E, 20$^\circ$-80$^\circ$N, has a small impact on the persistence of the regimes. 
It produces a longer persistence for NAO+, SB+ and AR+ and a shorter persistence for NAO-.

Figure \ref{fig:WR6sensdata} shows that, for the WR6 definition, extending the upper bound of the latitude to 90$^\circ$, has the largest effect on SB+ and SB-. 
This is to be expected, as the highs and lows of those regimes are located in the North. 
The region of 80$^\circ$W-30$^\circ$E, 20$^\circ$-80$^\circ$N, so a narrower region on the West, affects each regime differently, mostly decreasing the persistence. 
Extending the region to the East with the bounds 90$^\circ$W-40$^\circ$E, 20$^\circ$-80$^\circ$N, results in a lower occurrence for NAO- and a higher persistence for AR+, but has little other effects.

Figure \ref{fig:WR7sensdata} shows that for the WR7 definition, changing the region from the region used in~\textcite{Grams2017weather} to the region used in ~\textcite{vanderwiel2019regimes}, the largest effects are seen in the persistence. 
We see a large increase of the persistence for NAO- and EB+, and a decrease for SB+ and NAO+. 
For most regimes the occurrence is similar except for NAO+, where there is a large decrease and NAO-, where there is a large increase. 
For each regime the region of 90$^\circ$W-40$^\circ$E, 20$^\circ$-90$^\circ$N, closely resembles the results of the region used in~\textcite{vanderwiel2019regimes}. 
The region of 80$^\circ$W-30$^\circ$E, 20$^\circ$-80$^\circ$N, influences mostly the persistence. 
SB+ has a shorter and AR- has a longer persistence.
For other regimes, the results are similar to the results found by applying the reference settings. 
Investigating the results of the region 90$^\circ$W-40$^\circ$E, 20$^\circ$-80$^\circ$N, shows that results are similar to the results of using the region of ~\textcite{vanderwiel2019regimes}.
These combined results imply that for the WR7 definition, extending the region to 90$^\circ$W, has a significant impact.

The largest effects can be found in altering the period.
We have elaborated on these effects for the WR4 definition Section \ref{res:sensitivity} of the main text. 
There we also mention the effects of training the WR7 clustering algorithm on ERA5 reanalysis data from the period 1991--2020. 
The resulting regimes are shown in Figure \ref{fig:WR71991-2020pattern}. 
Here it is visible that the AR+ pattern is shifted Northward and the AR- regime lies above Great Britain, instead of the Atlantic Ocean. 
Furthermore, the high over Greenland of the NAO- regime is a smaller anomaly. 

\begin{figure}[hb]
    \centering
    \includegraphics[width=\linewidth]{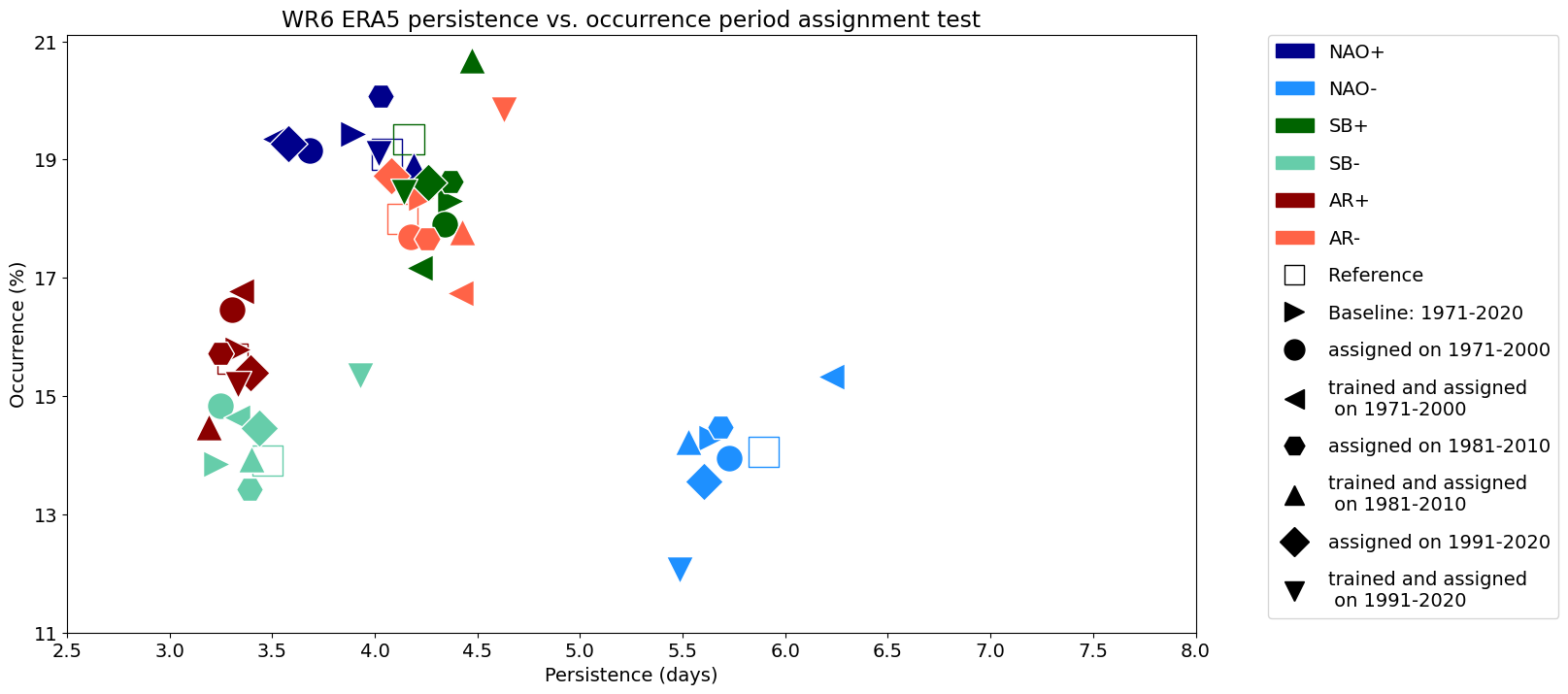}
    \caption{As Figure~\ref{fig:WR4periods}, but here the WR6 definition is applied to the different time periods. 
    All the expected regimes are found.
    The abbreviations of the regimes are explained in Section \ref{method:WRdef}.
    }
    \label{fig:WR6persoccperiod}
\end{figure}

\begin{figure}[p]
    \centering
    \includegraphics[width=0.85\linewidth]{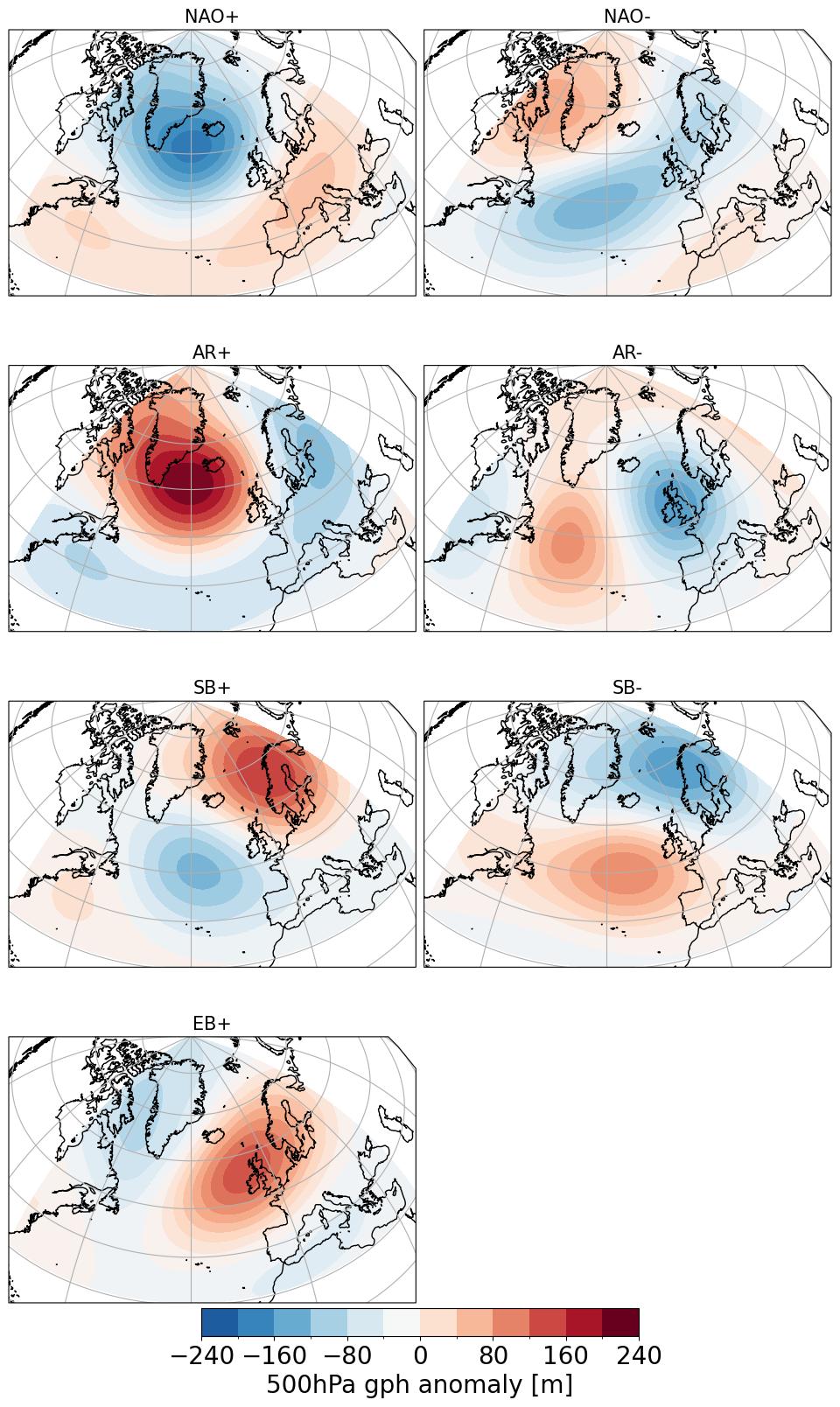}
    \caption{As Figure~\ref{fig:WRvisual}, but here the WR7 definition's clustering is done on the period 1991--2020, instead of 1979--2015.
    Though, all expected regimes can be identified, some are shifted, significantly when compared to their reference regime. For instance, here the maximum of the AR- regime lies above the United Kingdom, though originally it lies above the Atlantic Ocean. 
    }
    \label{fig:WR71991-2020pattern}
\end{figure}

Moreover, for WR6 we also investigated how the persistence and occurrence change if the data was clustered with the period 1971--2020, but assigned to a different period. 
The results of this experiment are shown in Figure \ref{fig:WR6persoccperiod}.
Here we can see a large spread in persistence and occurrence depending on which period the data was clustered with and which period the data was assigned to. 
This is not unexpected, because the clustering and assignment are dependent on trends within the data.

\end{appendices}

\end{document}